\documentclass[12pt,preprint]{aastex}

\def\tighttable{\def\baselinestretch{1.0}}
\def\arcsec{\ifmmode '' \else $''$\fi}
\def\arcmin{\ifmmode ' \else $'$\fi}
\def\arcsecpoint{\ifmmode ''\!. \else $''\!.$\fi}
\def\arcminpoint{\ifmmode '\!. \else $'\!.$\fi}
\def\cc{\ifmmode {\rm cm}^{-3} \else cm$^{-3}$\fi}
\def\cl{\ifmmode {\rm cm}^{-2} \else cm$^{-2}$\fi}
\def\micron{\ifmmode \mu{\rm m} \else $\mu$m\fi}
\def\kms{\ifmmode {\rm km\,s}^{-1} \else km\,s$^{-1}$\fi}
\def\Hubble{\ifmmode {\rm km\,s}^{-1}\,{\rm Mpc}^{-1}
	\else km\,s$^{-1}$\,Mpc$^{-1}$\fi}
\def\ergsec{\ifmmode {\rm ergs\;s}^{-1} \else ergs s$^{-1}$\fi}
\def\ergscm{\ifmmode {\rm ergs\,s}^{-1}\,{\rm cm}^{-2}
	  \else ergs\,s$^{-1}$\,cm$^{-2}$\fi}
\def\ergscmA{\ifmmode {\rm ergs\,s}^{-1}\,{\rm cm}^{-2}\,{\rm \AA}^{-1}
	  \else ergs\,s$^{-1}$\,cm$^{-2}$\,\AA$^{-1}$\fi}
\def\ergscmHz{\ifmmode {\rm ergs\,s}^{-1}\,{\rm cm}^{-2}\,{\rm Hz}^{-1}
	  \else ergs\,s$^{-1}$\,cm$^{-2}$\,Hz$^{-1}$\fi}
\def\Msun{\ifmmode M_{\odot} \else $M_{\odot}$\fi}
\def\Lsun{\ifmmode L_{\odot} \else $L_{\odot}$\fi}
\def\qo{\ifmmode q_{0} \else $q_{0}$\fi}
\def\Ho{\ifmmode H_{0} \else $H_{0}$\fi}

\def\gtsim{\raisebox{-.5 ex}{$\;\stackrel{>}{\sim}\;$}}
\def\tc{\tablenotemark{e}}
\def\ch{{\it Chandra}}
\def\HST{{\it HST}}
\def\ROSAT{{\it ROSAT}}
\newcommand {\lya}{Ly$\alpha$}

\shorttitle{Photometric redshift in CDFS}
\shortauthors{Zheng et al.}

\begin{document}

\title{Photometric Redshift of X-Ray Sources\\
in the Chandra Deep Field South\altaffilmark{1}}

\author{
W. Zheng\altaffilmark{2},
V. J. Mikles,\altaffilmark{2,3}, 
V. Mainieri\altaffilmark{4,5},
G. Hasinger\altaffilmark{5},
P. Rosati\altaffilmark{4},
C. Wolf\altaffilmark{6},\\
C. Norman\altaffilmark{2,7},
G. Szokoly\altaffilmark{5},
R. Gilli\altaffilmark{8},
P. Tozzi\altaffilmark{9}, 
J. X. Wang\altaffilmark{2},
A. Zirm\altaffilmark{10}, 
and
R. Giacconi\altaffilmark{11,2} 
}

\altaffiltext{1}{Based on observations performed at the European Southern 
 Observatory and observations with the NASA/ESA Hubble Space Telescope, 
obtained at the
Space Telescope Science Institute, which is operated by the Association of 
Universities of Research in Astronomy, Inc., under NASA contract NAS5-26555}
\altaffiltext{2}{Center for Astrophysical Sciences, Department of Physics and
 Astronomy, The Johns Hopkins University, Baltimore, MD 21218-2686, USA}
\altaffiltext{3}{Department of Astronomy, University of Florida, Gainesville,
 FL 32611, USA}
\altaffiltext{4}{European Southern Observatory,
 Karl-Schwarzschild-Strasse 2, D-85748 Garching, Germany}
\altaffiltext{5}{Max-Plank-Institut f\"ur Extraterrestrische Physik, 
 Giessenbachstrasse Postfach 1312, D-85741 Garching, Germany}
\altaffiltext{6}{Department of Physics, Denys Wilkinson Building, 
 University of Oxford, Keble Road, Oxford, OX1 3RH, United Kingdom}
\altaffiltext{7}{Space Telescope Science Institute, 3700
 San Martin Drive, Baltimore, MD 21218, USA}
\altaffiltext{8}{Osservatorio Astrofisico di Arcetri, Largo E. Fermi 5,
 50125 Firenze, Italy} 
\altaffiltext{9}{Osservatorio Astronomico di Trieste, via G. Tiepolo 11,
 34131 Trieste, Italy} 
\altaffiltext{10}{Department of Astronomy, Leiden Observatory, P. O. Box 9513, 2300 
 RA, Leiden, The Netherlands}
\altaffiltext{11}{Associated Universities Inc., 1400 16th Street NW, 
 Washington DC 20036, USA}

\begin{abstract}
Based on the photometry of 10 near-UV, optical, and near-infrared bands of the
Chandra Deep Field South, we estimate the photometric redshifts for 342 
X-ray sources, 
which constitute $\sim 99\%$ of all the detected X-ray sources in the field.
The models of spectral energy distribution are based on galaxies 
and a combination of power-law continuum and emission lines. 
Color information is useful for source classifications: Type-I AGN show 
non-thermal spectral features that are distinctive from galaxies and Type-II
AGN. The hardness ratio in X-ray and the X-ray-to-optical flux ratio are also 
useful discriminators. Using rudimentary color separation techniques, we are
able to further refine our photometric redshift estimations.
Among these sources, 137 have reliable spectroscopic redshifts, which we 
use to verify the accuracy of photometric redshifts and to modify the model
inputs. The average relative dispersion in redshift distribution is 
$\sim 8\%$, among the most accurate for photometric surveys. 
The high reliability of our results is attributable to the high quality and 
broad coverage of data as well as the applications of several independent 
methods and a careful evaluation of every source.
We apply our redshift estimations to study the effect of redshift on 
broadband colors and to study the redshift distribution of AGN. Our results 
show that both the hardness ratio and U-K color decline with redshift, which 
may be the result of a K-correction. 
The number of Type-II AGN declines significantly at $z>2$ and 
that of galaxies declines at $z>1$. However, the distribution of Type-I AGN 
exhibits less redshift dependence. As well, 
we observe a significant peak in the redshift distribution at z=0.6. 
We demonstrate that our photometric redshift estimation produces a 
reliable database for the study of X-ray luminosity of galaxies and AGN.

\end{abstract}


\section{INTRODUCTION}

The Chandra Deep Field South \citep[CDFS,][]{cdf1,cdf2,cdf5,1ms}, 
with approximately one million seconds of accumulated exposure, reaches a 
limiting flux of $5.5 \times 10^{-17}$ \ergscm\ in 
the 0.5-2 keV band and $4.5 \times 10^{-16}$ \ergscm\ in the 2-10 keV band. 
This is more than 20 times deeper than the \ROSAT\ deep 
survey \citep{rosat}. Within a $17\arcmin \times 17 \arcmin$ field, 
346 X-ray sources are identified.
CDFS has become one of best-studied fields in the sky, and its importance 
increases significantly with the GOODS observations \citep{goods} and the 
Hubble Ultra Deep Field \citep[UDF,][]{udf}. 

To reveal the nature and properties of these faint X-ray
sources, extensive follow-up observations have been made in many 
wavelength bands. Spectroscopic observations 
with the VLT and the FORS instrument \citep{fors} have yielded
168 spectra, of which 137 
have reliable redshifts. 
In the mean time, deep imaging observations have been made
with the ESO VLT and NTT in eight bands: U,B,V,R,I,J,H and K$_s$ 
\citep{eso}. 
The HST images taken by the GOODS team
provide high-quality photometry and source morphology. 
These observations reach significantly deeper than the spectroscopic 
observations, therefore yield an essential multi-band database for deriving
reliable photometric redshifts for the X-ray sources. They also provide two 
additional bands: the ACS/F850LP (z-band) and WFPC2/F300W (near-UV band). 

While X-ray sources account for only a small fraction of the field galaxies,
they form a special group with distinct properties. The \ROSAT\
Ultra Deep Survey \citep{lehmann} indicates that the majority of
X-ray sources are active galactic nuclei (AGN). Type-I AGN are characterized 
with broad emission lines and an underlying power-law 
continuum, and the Type-II AGN exhibit prominent narrow emission lines.
The combined results of CDFS and Chandra Deep Field North \citep{gilli2} 
further confirm that they are consistent with an assumption that obscured 
AGN (Type-II) outnumber normal, unobscured (Type-I) AGN by a factor of 
approximately 4-10.

A key parameter of these extremely faint X-ray sources is their redshifts, 
which may be determined spectroscopically or photometrically.
The spectral-energy-distribution (SED) method of redshift estimation 
\citep[photometric redshift,][]{koo,hdf,hyperz,bpz} uses 
a library of SED templates of galaxies at varying redshifts to fit the 
observed broadband data with a minimization of $\chi^2$.
The most prominent feature in these templates is a Balmer break around 4000 
\AA\ in the rest frame, because of the thermal nature of the galaxy continuum.
At high redshift ($z\gtsim 2$), intergalactic absorption shortward of \lya\ 
produces a significant break at $1216\ (1+z)$ \AA. These breaks or other 
strong spectral features can leave a distinctive signature across the range of 
passbands, making it possible to determine reasonable redshift estimates. 
Simulations and observations have demonstrated that photometric redshifts 
can be accurate on the order of 10\% for most sources.
The Sloan Digital Sky Survey (SDSS) uses five-band color selection
to find quasars, galaxies and other sources \citep{sdss2,sdss3}. Photometric 
redshift estimation is an extension of such color selection techniques, and 
it has gained popularity in recent years as deep imaging 
surveys become one of the most important methods in probing the early universe
\citep{eis3,wolf}. The majority of sources in these fields are so faint that it 
is impractical to obtain their spectra in large number.  
Redshift information
used in conjunction with the optical, UV, IR, and X-ray photometry is central
in determining the nature of faint X-ray sources and how they are
linked to the galaxy population at similar redshift. Also, we can
explore how the luminosity function of X-ray galaxies differs from that
of other galaxies. A number of studies have been carried out to compare
photometric and spectroscopic redshifts \citep{gonzalez,mobasher,gandhi} for 
X-ray sources. A study of similar scope \citep{barger} finds 
that the photometric redshifts are within 25\% of the spectroscopic redshifts 
for 94\% of the non-broad-line sources with both photometric and spectroscopic 
measurements.

In this paper, we focus on refining our photometric redshift techniques on an 
X-ray selected sample of sources
and testing our results against current model predictions.
Our redshift estimates are based on public imaging data
in the near-UV (NUV, ~3000\AA), optical and infrared bands.
The results will facilitate our understanding of the broadband properties, 
luminosity function of X-ray sources and many other important issues such as 
clustering and the large-scale structure of the early universe.

We discuss the data resources and processing in \S2. In \S3 we discuss
the application of several methods of photometric redshift and compare their 
results to derive the most probable redshifts. With the photometric redshifts 
for all CDFS sources, we discuss in \S4 the redshift dependence of source 
properties and potential implications.

\section{DATA}

\subsection{X-Ray Data}

Our sample is the 346 X-ray sources identified from the 940 ksec of CDFS data 
\citep{1ms}. 
Throughout this paper we use AB magnitudes, and we calculate the 
X-ray magnitudes as $X = -2.5 \log f + 2.5 \log \nu -48.6$, 
where $f$ is flux in units of \ergscm, and $\nu = 2.4 \times 10^{17}$ 
and $9.5 \times 10^{17}$ Hz for the soft X-ray band (0.5-2 keV) and hard 
X-ray band (2-10 keV), respectively. 
For sources not detected in one of the X-ray bands, we derive the upper 
limits to X-ray magnitudes. In this calculation, we consider the limiting 
magnitude for the soft and hard-X-ray as $\sim 35.5$ ($f \sim 5.5 \times 
10^{-17}$ \ergscm) and $\sim 34.7$ ($4.5 \times 10^{-16}$\ergscm), 
respectively. 
Approximately 85\% of these X-ray sources have been identified in the optical
R-band images of ESO VLT/FORS \citep{1ms}.

\subsection{Imaging Data} 

The ESO/FORS R-band images cover most of the CDFS field to a limiting magnitude
of $\sim 26.5$. Supplementary positions are added from the comparison with the 
ESO Imaging Survey \citep[EIS,][]{eis3}. 
The EIS images and catalogues also provide multiband information.
Only objects that fall with 5\arcsec\ from the X-ray positions are selected.
Each entry in the catalogs is compared with the X-ray positions \citep{1ms}.
The $5 \sigma$ limiting magnitudes of the EIS data are 
U=25.7, B=26.4, V=25.4, R=25.5, I=24.7, J=23.4, and $\rm K_s=22.6$.
The VLT/GOODS imaging data in J, H, and K$_s$ bands  reach limiting 
magnitudes of $\sim 25.5$.

Uncertainties of the 
optical zero point are in the range of 0.03 - 0.08. As our comparison of 
the measurements from different data indicates, real uncertainties may be 
larger than tabulated values. We therefore assign uniform errors of  
0.1 magnitudes for most data points unless the tabulated errors are larger 
than 0.1. 
As shown in Figure 1 and Table 1, the majority of the optical counterparts lie 
with 2\arcsec\ from the X-ray position, (and within 1\arcsec\ from the 
optical position given in \citet{1ms}).

Additional data are obtained from the COMBO-17 survey 
\citep{combo,combo2,combo3} in its most sensitive broad bands
and from the GOODS images. The COMBO-17 data of the CDFS consist of catalogs 
in three broad bands (B, V and R), with limiting magnitudes of $\sim 26$. 
The photometric redshift of 272 matched X-ray sources are available. 
In COMBO-17 every object is either classified as a star, a galaxy or a 
quasar, and in the latter two cases redshifts are determined.
COMBO-17, because its significantly higher number of bands, provides more 
reliable redshift estimations, particularly for those with emission 
lines. A comparison between the COMBO-17 results and spectroscopic redshift
yields an average error of $1-2\%$ \citep{combo3}, better than that of our 
broad-band results ($\sim 8\%$). However, its advantage is limited to 
relatively bright sources ($R<24$) and redshift below 1.5 or above 2.5. 

The optical GOODS data \citep{goods2} are taken with the Advanced Camera 
for Surveys (ACS) on board of \HST\ and filters F435W, F606W, F775W, and 
F850LP (B,V,I, and Z band, respectively). These data reach magnitudes of 
$\sim 28$, therefore providing additional significant resources. 
The processed ACS data images and catalogs, including the UDF, are 
retrieved from the Multimission Archive at the Space Telescope Science 
Institute (MAST). We use the 
{\it SExtractor} algorithm \citep{sex} to identify sources. 
Since the HST data have significantly higher spatial resolution, we sum all 
the source fluxes within 0.\arcsec 5 from the optical positions in the EIS 
data, in order to match its photometry. Additional source positions are 
obtained for those that are not detected in either the EIS or FORS-R data. 
In total, data of 69 sources are added to the EIS database.

To obtain reliable colors for our sources the spread in seeing conditions for 
images in different wavebands has to be taken into account. We have 
computed corrected magnitudes according to the following strategy: 1) We 
computed magnitudes in apertures of 2\arcsec\ in each waveband using 
the available images ("original" images); 2) We degraded the point 
spread function (PSF) of each image to match the worst condition 
("degraded" images); 3) We recomputed the magnitudes in apertures of 2 
\arcsec\ using the "degraded" images; and 4) We derived corrections for 
the different seeing conditions comparing, for stellar "bright" objects, 
magnitudes in the original and in the "degraded" images and applied 
these corrections to the "original" magnitudes.

The GOODS data also include the band of WFPC2 F300W, which provides clues 
that complement the optical data. In the low-redshift range, these data 
enable the separation of power-law continua from thermal components. At 
$ z > 1.5$, the data may reveal the signature of prominent \lya\ 
emission. We obtain the processed images from MAST and use IRAF tasks to
reject cosmic ray events and stack the images. 
A total of 112 stacked field images are produced, and each of them contains 
at least three independent WFPC2 images. We use {\it SExtractor} to identify 
sources from the 
stacked images. The limiting magnitude in this band varies with the number of 
stacked images. The average $5 \sigma$ limiting magnitude is $\sim 26$. 
105 sources have detection in the NUV band, and 123 are confirmed with no 
detection.

In all, 191 sources have data in all ten bands: NUV,
U, B, V, R, I, Z, J, H, and K$_s$.
326 sources 
have data in seven or more broad bands between U and K$_s$, 15 other sources 
have 
five or six bands of data. 14 sources have no infrared data. There are four 
objects which do not have detections in addition to the FORS-R results 
\citep{1ms}. 

\subsection{Spectroscopic Data}

Spectroscopic observations of selected X-ray sources were carried out in 2000 
and 2001, with the VLT telescope and FORS instrument. Details of the 
observations and data reduction may be found in \citet{fors}. 
The spectroscopic redshifts for 137 objects provide a critical source for our 
verifications, and the data also allow us to distinguish between 
broad-line and narrow-line objects. 
Because of the low spectroscopic resolution ($R \sim 300$), the 
classification of sources is tentative. As defined in \citet{fors}, 
objects with emission lines of $>2000$ \kms\ are considered as 
Type-I AGN. 
Such a definition based on the opticaal spectroscopy is largely consistent 
with the classification by X-ray data (see \S 3.2 and Fig. 3), but a few 
exceptions do exist.
Approximately twenty spectroscopic redshifts are derived from the sources
fainter than the nominal limiting magnitudes of $R \sim 24$. These values 
are considered reliable because of the sources' prominent emission-line 
features.

\section{MODELING}

\subsection{Photo-z Codes and Galaxy Templates}
We calculate photometric redshifts using two parallel models: HyperZ
\citep{hyperz} and the Bayesian Model \citep[BPZ,][]{bpz}. 
The results are dependent on the selection of model templates.
Using spectroscopic redshifts to check the reliability of 
the derived photometric redshifts, we are able to choose the galaxy 
templates that best suit. For HyperZ, our sample of X-ray sources
include starburst galaxies, E, S0, Sbc, Scd, 
and Im galaxies from \citet{cww} and the GISSEL98 library \citep{bc}.
The BPZ model combines the $\chi^2$ minimization and Bayesian marginalization,
using relevant knowledge, such as the expected shape of the redshift
distributions and the galaxy type fractions. We use the default library in the
BPZ, which is similar to HyperZ but includes additional templates of 
starburst galaxies, and a set of flat spectra. 

Both the HyperZ and BPZ models provide accurate fits for approximately
85\% of the objects, however, there are a number of objects 
whose fitted redshifts are different from the known spectroscopic values.
Most of these sources are Type-I AGN, which are known for their power-law 
spectra with prominent broad emission lines. Failure to detect strong broad
lines in the passbands is the most likely cause for poor fits.
Even in the broadband data, their difference may be significant. In 
Figure 2 we plot three representative sets of broadband data, plus
the SDSS composite spectrum as a representation of Type-I AGN and
quasars \citep{sdss1}.

Between approximately 1200 and 5000 \AA\ the underlying 
continuum may be described by a single power law. However, the SDSS composite 
spectrum itself does not fit all the Type-I AGN. Some Type-I AGN show 
relatively weak emission lines, and their continuum is dominated by a thermal 
continuum of the underlying galaxy, instead of a power law.
We therefore generate a set of AGN spectra by combining a continuum component
and emission-line components. The continuum is a broken power law. 
We assume a power law of $f_\nu \propto \nu^{-\alpha}$, with $\alpha$ 
varying between 0 and 1. Longward of 5000 \AA\ we add another power-law 
component of $\alpha = 1.4$ to represent the contribution of the
host galaxy, and a thermal component in the infrared band \citep{malkan,elvis}.
The line components have the same ratios as those derived 
in \citet{francis}, but the line intensities vary. 
The \lya\ equivalent (and other lines accordingly) is scaled by factors 
between 0.5 and 5.0 from their nominal values in the composite quasar spectrum
\citep{francis}. The H$\alpha$ intensity is derived from the 
H$\beta$ value, scaled up by a factor of three.  

\subsection{Color Separation}

Running photometric-redshift tasks with two sets of templates of galaxies and 
power-laws, we test the results with objects with known spectroscopic redshifts.
While HyperZ can pickup the majority of Type-I objects by the lower $\chi^2$ 
values in the fitting output, it identifies $\sim 20\%$ of them as galaxies. 
Future work of improving the power-law templates and/or selective choice of 
galaxy templates may be needed. However, this uncertainty may be alleviated 
using the additional information in the X-ray bands. 
As discussed in \citet{fors}, the X-ray properties are indicative of object 
types. We pre-divide our sample into three groups based on their X-ray properties:

\begin{itemize}
\item Type-I AGN: $\rm X_S - X_H > 0.1$ or and $\rm X_H - R < 12$. 
 Power-law templates;
\item Type-II AGN: $\rm X_S - X_H < 0.1$. Galaxy templates;
\item Galaxies: $\rm X_S - X_H > 0.1$ and $\rm X_H - R > 12$. 
 Galaxy templates.
\end{itemize}
The groups are established primarily by their X-ray 
hardness ratio and the X-ray-optical flux ratio, not on the broadband 
optical and infrared data.
As shown in Figure 3$B$ and 3$D$, nearly all Type-II AGN have X-ray magnitude 
difference $\rm X_S - X_H > 0.1$. 
This parameter can also be interpreted as the hardness ratio 
(H-S)/(H+S) $> -0.2$, where H and S are the counts in 
the hard (2-10 keV) and soft (0.5-2 keV) bands, respectively.  
In addition, most galaxies are weaker emitter in the hard X-ray band, 
at $\rm X_H - R > 12$, where $R$ is the optical R-band magnitude. Once the 
initial separations are made, 
we manually check the broadband color of each object.

These selections are applied before our photometric-redshift estimations. 
The estimations for Type-II AGN and galaxies are made using the
templates in HyperZ and BPZ, as their continuum features are primarily 
thermal. For Type-I AGN, we use our power-law templates and the HyperZ task,
which allows a large number of templates. To check whether the results is 
overly dominated by one single band, we also run the datasets with and without 
the bluest band. This is one of the criteria used in the final 
determination of redshifts.

The source classification in Table 1 represents our best results: For the 137 
objects where reliable redshifts are established, we use their broad-line 
properties to classify them. For the others, we use the COMBO-17 
results. If sources are not in the COMBO-17 catalog, we use their X-ray 
properties to classify them. In a few cases when we have reasons to believe 
as exceptions, we discuss them in \S 4.3.

\section{DISCUSSIONS}

There are several sources of uncertainties and discrepancies in our redshift 
estimations: (1) Source confusion: Deep optical images often reveal multiple 
sources within the X-ray positional error circle. Many factors, such as the X-ray 
brightness and optical morphology, need to be considered before a careful 
decision may be made. (2) Complicated source energy distributions: Some X-ray 
sources have hybrid continuum shape, as the combination of a thermal component 
and a power law. Our model templates may not be sophisticated enough to reach 
the best and unique fits; and (3) Faintness of some sources: When the magnitudes
are close to the detection limit, the task do poorly in distinguishing the 
Balmer and Lyman breaks. 

\subsection{Photo-z Confidence Levels}

The results are cross-compared between HyperZ and BPZ. Our main concern is potential 
catastrophic errors between redshift $\sim 0.5$ and that of $\sim 2.5$, which 
arise from the confusion between the Balmer continuum break and 
Lyman-$\alpha$ break. 
In such cases, the X-ray properties provide additional discriminators for the 
estimates of redshift.
As shown in Figure 4, approximately 90\% of the redshift values derived from 
BPZ and HyperZ match well. Their photometric redshifts and errors are calculated as 
the weighted average of BPZ and HyperZ. The weights are taken as the reciprocal 
of redshift errors, averaged over the lower and higher sides. Both BPZ and 
HyperZ output the confidence levels of photometric redshifts, which we use to 
calculate the errors in redshift. The new 95\% confidence level of photometric redshift
is calculated from the weighted average of errors, respectively, 
on the lower and higher redshift side. With such a method, the redshift values 
with narrower redshift range are given with more weights. Several pairs of
seemingly discrepant redshifts are compromised as they converge towards the 
values with narrower redshift ranges. 
Several COMBO-17 redshifts do not have published error values \citep{combo3}.
We use the errors in the pairing BPZ or HPZ redshifts.
Discrepant redshifts in 
approximately 15 sources are likely the results from (1) faint sources for 
which the uncertainties rise; and (2) some sources with power-law continua, for 
which the HyperZ evaluation is more reliable because of the added AGN templates.
Nearly all the hard X-ray sources as well as the objects with large R-K colors 
are at lower redshifts ($z < 1.5$). Figure 5 compares 
photometric and spectroscopic redshifts. The average dispersion 
$\Delta z /(1+z)$ is 0.08. We run a similar test for BPZ and HyperZ values. The
dispersion is 0.08 for HyperZ values, and 0.10 for BPZ.

We assign quality indices to every source, from 0 to 3. With quality index 3,
the spectroscopic redshift is certain, and so is the optical counterpart.
It corresponds to quality index 2+ and higher in \citet{fors}.
Index 2 means a secure spectroscopic redshift, but the optical counterpart is
uncertain.
Other index values are assigned with multiple factors, such as the COMBO-17 
(0.4), BPZ (0.3), HyperZ (0.2), and single-line redshifts (1.0). 
The index values of 1 or smaller can be additive, e.g. if we have a 
single-line spectroscopic redshift, for which all three photo-z results agree,
Q=1.9. 
The grading in photo-z does not mean to indicate a quality
flag between BPZ and HyperZ. 

Our main results are presented in Table 1. 
Three objects are marked with quality index of 0.2. In two such cases, a 
power-law continuum is apparent, and their fitting results with HyperZ must be
specially chosen. In another case, the HyperZ values are supported by the 
X-ray spectroscopy. 
Two objects are marked with quality index of 0.3. It is very faint and 
are detected only in a few bands, leading to confusion between \lya\ and Balmer
breaks. In such cases, the BPZ values are chosen.
Based on the redshift information, we are able to fine tune the 
classification in \S3.2 and further divide the three groups into seven:
 
\begin{itemize}
\item Galaxy: $L_x < 10^{42}$ \ergsec\ with hardness ratio $HR < -0.2$. 
\item Type-2 AGN: $10^{42}< L_x < 10^{44}$ \ergsec\ and the hardness ratio 
$HR > -0.2$. 
\item Type-2 QSO: $L_x > 10^{44}$ \ergsec\ and the hardness ratio $HR >
 -0.2$. 
\item Type-1 AGN: $ 10^{42}< L_x < 10^{44}$ \ergsec\ and the hardness ratio
 $HR < -0.2$. 
\item Type-1 QSO: $L_x > 10^{44}$ \ergsec\ and the hardness ratio $HR <
 -0.2$. 
\item M-star
\end{itemize}

Figure 6 displays a comparison between
the COMBO-17 results and our photometric redshifts for sources which do not 
have reliable spectroscopic redshifts. The latter are derived from the 
combination of BPZ and HyperZ, as described above.
The redshift dispersion between these two samples is approximately 12\%.
\subsection{Comments on Individual Objects}

The high quality of the \HST\ images enables us to resolve many X-ray sources.
We find that approximately forty sources show significant sub-arcsecond
structure, {\it i.e.} two or more components with flux contrast less than 10.
They are marked in Table 1 (in column "Offset").
Separation beyond that would have been identified in the EIS data.
Three such sources show considerable color difference between bands, 
therefore implying discrepant redshifts. 

Several sources show sub-arcsecond components of different colors. 
The source numbers refer to that in \citet{1ms}, representing a 
unique detection number for each source. 
The optical cutout images for four such sources are displayed in Figure 7.
 
CDFS 27: This source consists of three components. 
Component A exhibits a similar magnitude to the other two, but it is 
significantly fainter in other bands.

CDFS 55:  As shown in Figure 7, this source consists three distinct components
in the V and B band separated by $\sim 0.\arcsec 25$, and is extended in all 
four ACS bands. 
The component A is significant fainter in the B band, therefore implying 
a possibly different redshift. 

CDFS 94: Three components within the X-ray error range. 
They can be seen marginally in Figure 13 of \citet{1ms}.

CDFS 630: Multiple sub-arcsecond structure are shown at the bottom of Figure 
7. Component B is brighter in the Z band. 

The properties of a number of peculiar sources also need some discussions:
are discussed in \citet{vinc}. Here we discuss several other sources:
 
CDFS 38 and 100: The broad-band data of these two objects exhibit a 
power-law continuum without 
signs of emission lines.  The spectrum of source 100 shows a single emission 
line around 8600\AA, which may be O [{\sc ii}] 3737 at redshift 1.309.

CDFS 6 and 626: The data resemble a power-law continuum without a 
significant break. The HyperZ values may be more accurate. Alternative values
of redshifts, based on galaxy models, are given near the end of Table 1.

CDFS 217, 243 and 508: Their primary BPZ values are at $z>7$, but the X-ray 
spectroscopy reveals some features, which resemble iron K$\alpha$ lines. If 
we take the HyperZ values or the secondary BPZ values, the redshifts are 
between 2.5 and 4.6, are consistent with the X-ray results. More details will 
are discussed in \citep{vinc}.
 
CDFS 523: This source was within 2\arcsec\ from the list of red objects
by \citet{koek}. 
Since no confirmed detection is given in any optical or infrared bands, the 
counterpart may be another one which lies 1.9\arcsec\ and is weakly detected 
in several HST/ACS bands (AB$ \sim 27.4$).

Table 2 lists four sources that are weakly detected only in one band 
\citep[FORS-$R$,][]{1ms}, but not confirmed in the EIS multi-band images:

CDFS 261: $R-$band magnitude $> 26.1$. No \HST\ data are available. 

CDFS 616: $R-$band magnitude $> 26.1$. No \HST\ data are available. 

CDFS 640: $R-$band magnitude $> 25.7$. No \HST\ data are available. 

CDFS 649: $R-$band magnitude $> 25.7$. But the source detection is not 
confirmed by the \HST\ data. The nearest counterpart in the \HST\
 data is 14\arcsec\ away.  

No redshift information can be derived from these single-band detections.
Furthermore, these sources are among the faintest in both the X-ray and 
optical bands. It is therefore possible that their detections may not be 
highly reliable.

\subsection{Effect of Filter Bands and Sensitivity}

Approximately 95\% of the X-ray sources have infrared data. To test their effect 
on photometric redshifts, we run the tasks with only the optical bands. The 
average accuracy in redshift is reduced from $\sim 8\%$ to $\sim 12\%$. But the 
effect is redshift dependent: The most sensitive redshift ranges are between 
$0.8 < z < 2.5$, where the accuracy is at a level of 20\%. This region is known 
for the lack of prominent spectral features in the optical bands. At other 
redshifts, the impact of infrared data is not so significant.

Our photometric redshifts also depend on the faintness of sources. If we 
establish a subgroup whose R-band magnitudes are within 1.5 magnitudes to the 
detection limit, the number of outliers increase significantly. This dependence 
on sensitivity may be from the same reason as stated above: Near the detection 
limit, the tasks may not distinguish Balmer breaks from \lya\ breaks. 

As discussed in \S 4.1, the dispersion of HyperZ results, as compared with 
spectroscopic redshifts, is slightly smaller than that for BPZ, This may be 
understood as our HyperZ fitting includes a large grid of simulated AGN-1 
spectra, while the BPZ fitting uses only its built-in galaxy templates.
Most spectroscopic redshifts are obtained for relatively bright 
sources, which include a large portion of Type-I AGN and QSOs. For fainter 
sources, we believe that BPZ has advantages as it takes the priors into 
consideration.

\subsection{Evolution of X-Ray Properties}

Figure 8 displays the redshift distribution of X-ray source numbers for the 
three groups defined in \S3.2.
Type-II AGN are mostly populated at $z<1.5$, and galaxies at $z<1$. Type-I AGN
exhibit less redshift dependence. We also observe a significant peak in the 
redshift distribution at $z \sim 0.6$, which
confirms a report \citep{cdf2,fors} 
on the large-scale structure at several redshift positions.

Figure 9 displays the redshift dependence of X-ray luminosity.
The X-ray luminosity is calculated with parameter $H_0 = 71$ \Hubble,
$\Omega_m=0.3$ and $\Omega_\lambda=0.7$. The intrinsic luminosity of galaxies
forms a low-end envelope. 
The hard-X-ray luminosities of Type-II AGN are comparable to that 
of Type-I AGN, but different in the optical. This may be in agreement with 
the assumption that they are intrinsically similar, but viewed at 
different angles. At edge-on positions, the optical and soft-X-ray flux may be
subject to significant obscuration.

Our results are consistent with that derived from the spectroscopic redshifts
\citep[Fig. 22,][]{fors}, offering encouraging evidence that, with careful 
calibration, photometric redshifts are a reliable resource for the statistical
study of distant sources. The redshift values in this paper have found their
application in a recent study of the X-ray luminosity function \citep{xlf}.

\subsection{Redshift Dependence of Color}

As shown in Figure 10, there is significant evolution with redshift
of X-ray and optical/near-IR colors. The hardness ratio declines with redshift.
This is consistent with the finding of \citet{gilli}
that most hard X-ray sources are at redshift lower than 1.5.
While it is possible that a part of the observed trend may be the result of 
increasing luminosity threshold with redshift, the effect may not account for 
all the changes, as they are found in sources whose fluxes are well above the 
detection thresholds. At higher redshifts, a part of the hard Xray flux is 
shifted into the soft X-ray band, thus reducing the hardness ratio. This
effect may also explain the change in the U-K color: At higher redshifts, the 
K-band data represent only the data at the restframe wavelength $2.2/(1+z)\ 
\mu$m, since the reddening effect becomes significantly lessened. 

The change in the U-K color is mainly due to a simple redshift effect: At 
higher redshifts, the K-band data represent only the data at the restframe 
wavelength $2.2/(1+z)\ \mu$m, since the reddening effect becomes 
significantly lessened. 

As discussed by \citet{fors}, different types of X-ray sources are separated 
in the plot of X-ray hardness ratio vs. X-ray luminosity. Figure 11
demonstrates the same trend, with a significantly higher number of sources. We
plot the relationship for both the hard- and soft-X-ray luminosities. Many 
galaxies are not detected in the hard X-ray band.

\section{SUMMARY}

We utilize a photometric redshift estimation using 12-band data in 
near-UV, optical, infrared and X-ray. We employ a set of power-law 
models for Type-I AGN and various models of galaxies, to
derive the photometric redshifts for 342 CDFS sources.
A comparison with the spectroscopic redshift of 137 objects suggests an 
average accuracy of approximately 8\%. 
Our results well match that of COMBO-17. 
These encouraging results suggest that photometric redshifts are of 
significant accuracy. 

The redshifts of CDFS sources subtend a broad range between 0.01 to 4.66. 
The hard X-ray sources, mostly Type-II AGN, are mainly populated at $z<1.5$. 
The sources with power-law continuum, {\it i.e.} Type-I AGN, show 
less redshift dependence. 

\acknowledgments

Observations have been carried out using the ESO New Technology Telescope 
(NTT) and the 2.2m telescope at the La Silla observatory under Program-ID Nos.
61.A-9005(A), 162.O-0917, 163.O-0740, and 164.O-0561. 
This work has benefited from the GOODS project, which provides high quality
VLT and \HST\ images and source catalogs for a significant portion of the 
CDFS field. The HST data presented in this paper were obtained from MAST. STScI 
is operated by the Association of Universities for Research in Astronomy,
Inc., under NASA contract NAS5-26555. Support for MAST for non-HST data is 
provided by the NASA Office of Space Science via grant NAG5-7584 and by other 
grants and contracts.

This work has been supported in part by NASA grant NAG-8-1527 and NAG-8-1133.
W.Z. thanks Roser Pell\'o and Narciso Ben\'{\i}tez for their kind help and advice on
photometric redshift estimations.

\clearpage

\centerline{\bf Figure Captions}
\bigskip

\figcaption{Position offsets between the X-ray and Optical/Infrared data.
Some sources are not detected in the EIS survey, and supplementary data 
are collected from the \HST\ imaging data.
\label{fig1}}

\figcaption{Multiband color of selected CDFS sources in restframe.
The dotted lines 
represent Type-I quasars; 
dashed lines - Type-II quasars; and solid 
lines - galaxies. A composite SDSS spectrum of quasars (Vanden Berk et al.\ 
2001) and a power-law template (without Ly$\alpha$ forest absorption) are 
plotted for comparison.
\label{fig2}}

\figcaption{Properties of CDFS sources with known spectral type and redshift:
(A) X-ray-to-optical ratio vs. redshift; (B) Hardness ratio vs. redshift; 
(C) X-ray-to-optical ratio vs. optical/infrared color; and (D) 
Hardness ratio vs. optical/infrared color. 
The open circles represent Type-I AGN; open squares Type-I quasars;
filled circles Type-II AGN; filled squares Type-II quasars; triangles 
galaxies, and asterisks clusters.
\label{fig3}}

\figcaption{Comparison of photometric redshifts of BPZ and HyperZ.
The symbols are the same as Figure 3. 
Objects fainter than $R<25$ are marked with smaller symbols.
Stars represent the cases where broad-band SED resemble a power law.
In most of these cases, spectroscopy or COMBO-17 data confirm the HyperZ
values derived from power-law templates.
\label{fig4}}

\figcaption{Comparison of photometric redshifts with spectroscopic
redshift. The photometric redshifts are derived from Hyper-Z and BPZ.
The symbols are the same as Figure 3.
The object marked with an arrow 
consists of components of significantly different color (CDFS 55).
\label{fig5}}

\figcaption{
Comparison of photometric redshifts with COMBO-17 results. 
The photometric redshifts are derived from Hyper-Z and BPZ. 
The selection criteria are: (1) COMBO-17 magnitude is brighter than 24 in 
the $R$ band; and (2) Spectroscopic redshift is unknown.  
The symbols are the same as Figure 3. 
\label{fig6}}

\figcaption{ACS cutout images of four sources with sub-arcsecond components. 
Their CDFS source identification numbers are marked at upper-left corner of each image, 
and the optical band name at the lower-right corner. The circles represent the 3$\sigma$
 X-ray positional errors, along with intensity contours. The dimension for 
each cutout is 3\arcsec, and the orientation follows that of the relevant ACS 
images, not aligned with the east-north direction. See discussion in \S 4.2.
\label{fig7}}

\figcaption{Redshift distribution of X-ray sources. The shaded 
histograms show that for the sources with spectroscopic redshifts.  
Galaxies are detected up to $z\sim 2$. They do not show an excess at 
$z\sim 0.7$.
\label{fig8}}

\figcaption{X-ray luminosity vs. redshifts. 
The symbols are the same as in Figure 4.
The curve represents the boundary for limiting flux.
The symbols are the same as Figure 3.
\label{fig9}}

\figcaption{Redshift dependence of optical-infrared color and X-ray 
hardness. The symbols are the same as Figure 3.
\label{fig10}}

\figcaption{X-ray hardness vs. X-ray luminosity. The symbols are the 
same as Figure 3.
\label{fig11}}
\clearpage

\setcounter{figure}{0}
\begin{figure}
\plotone{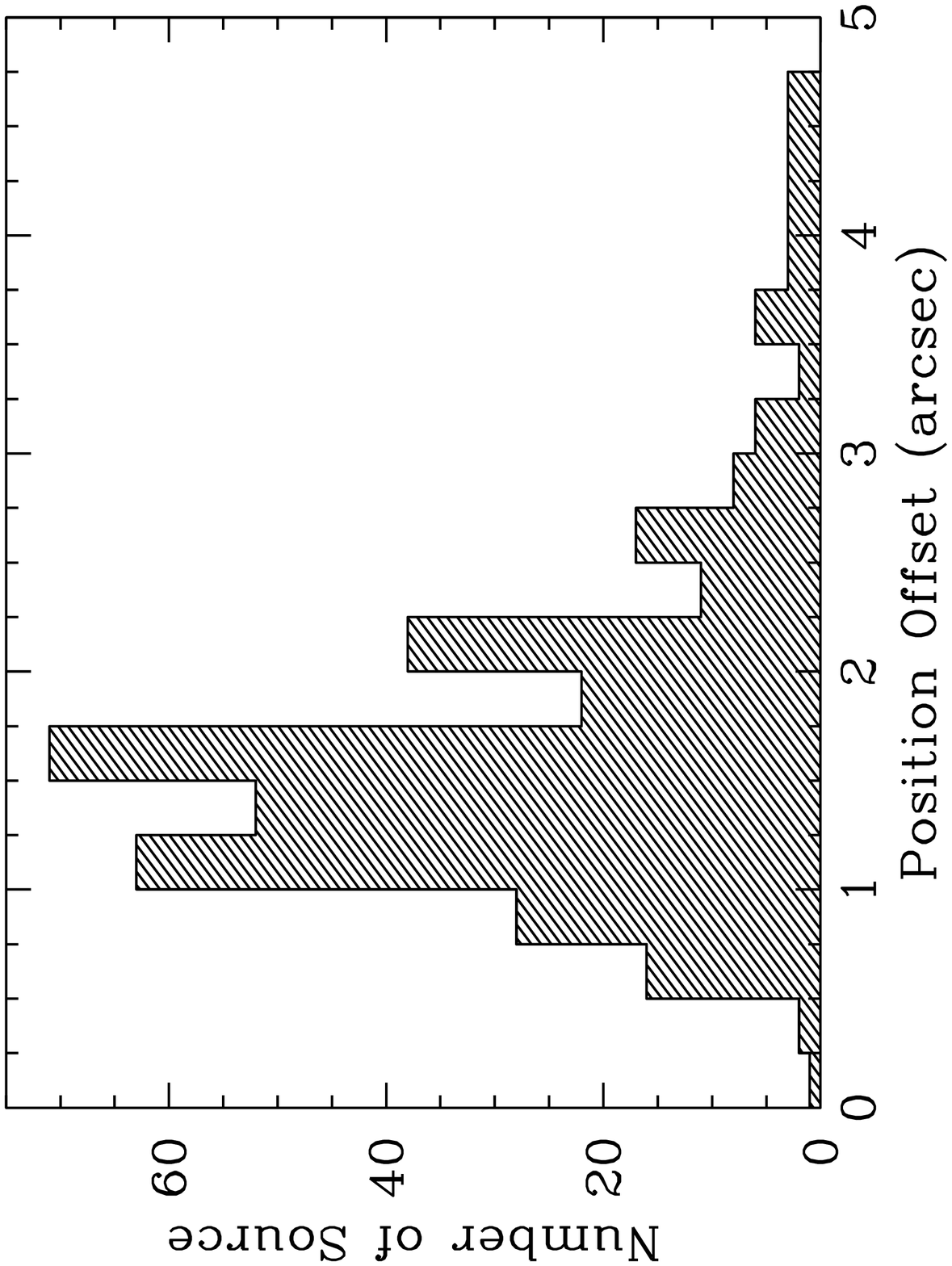}
\caption{~}
\end{figure}

\begin{figure}
\plotone{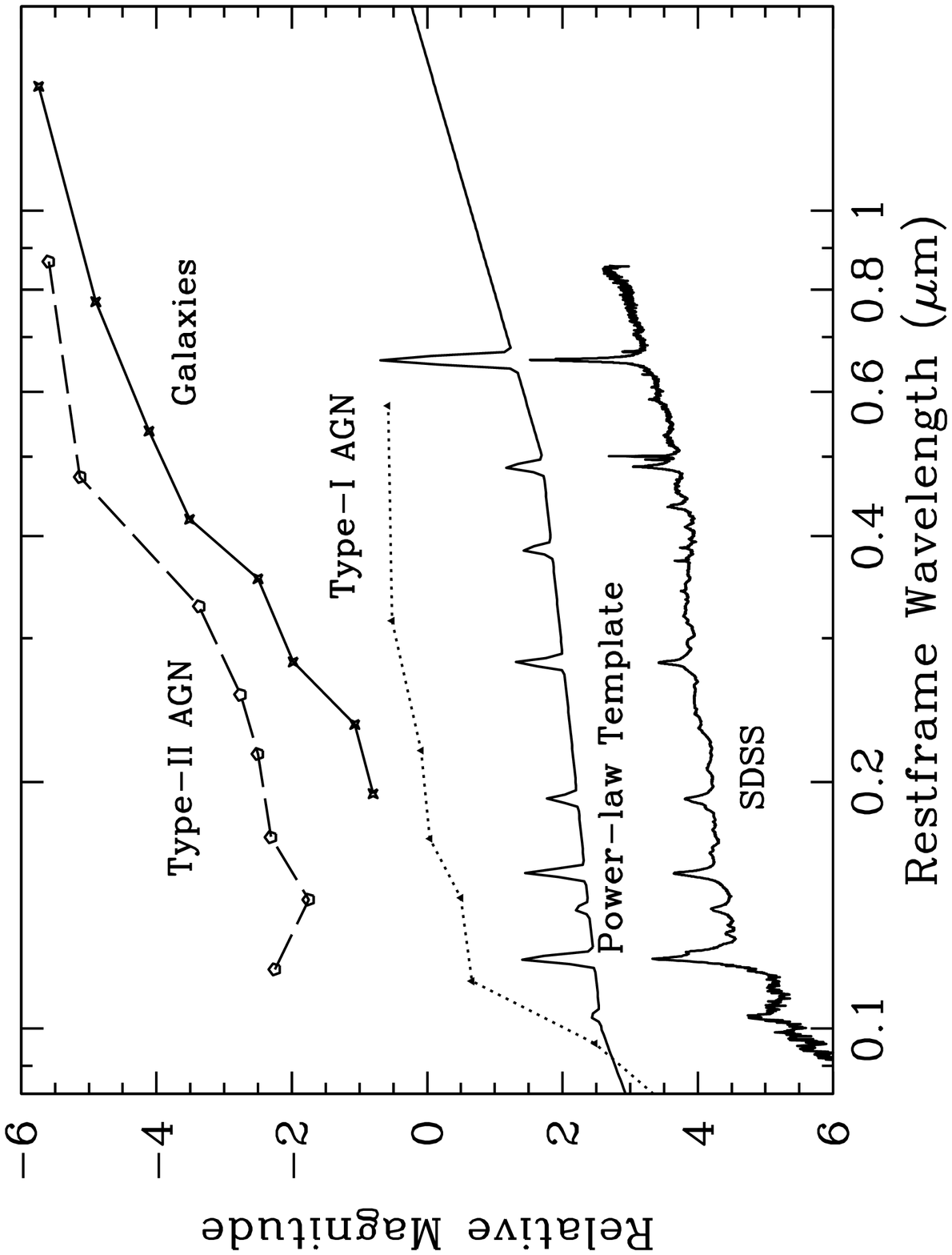}
\caption{~}
\end{figure}

\begin{figure}
\plotone{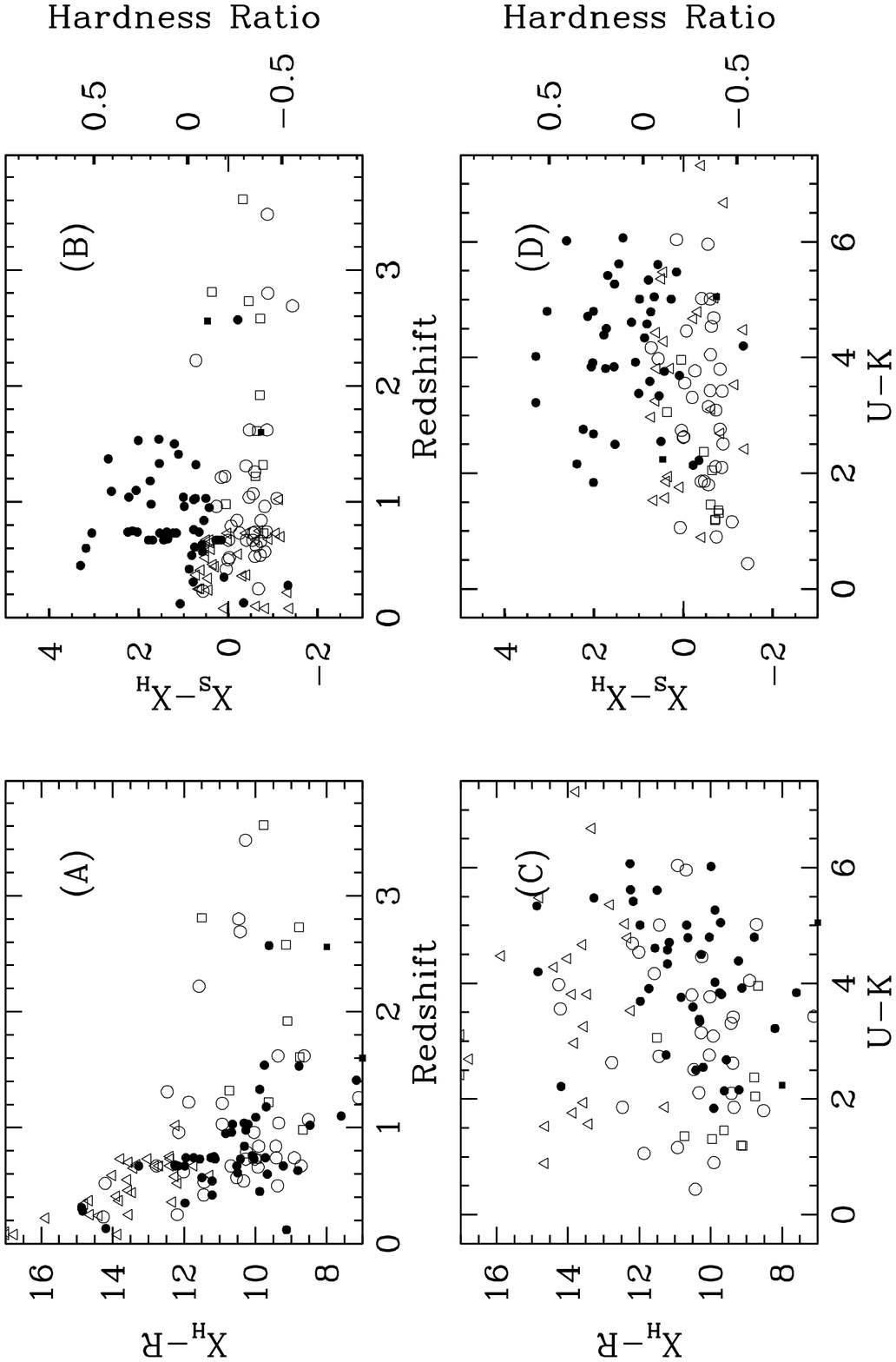}
\caption{~}
\end{figure}

\begin{figure}
\plotone{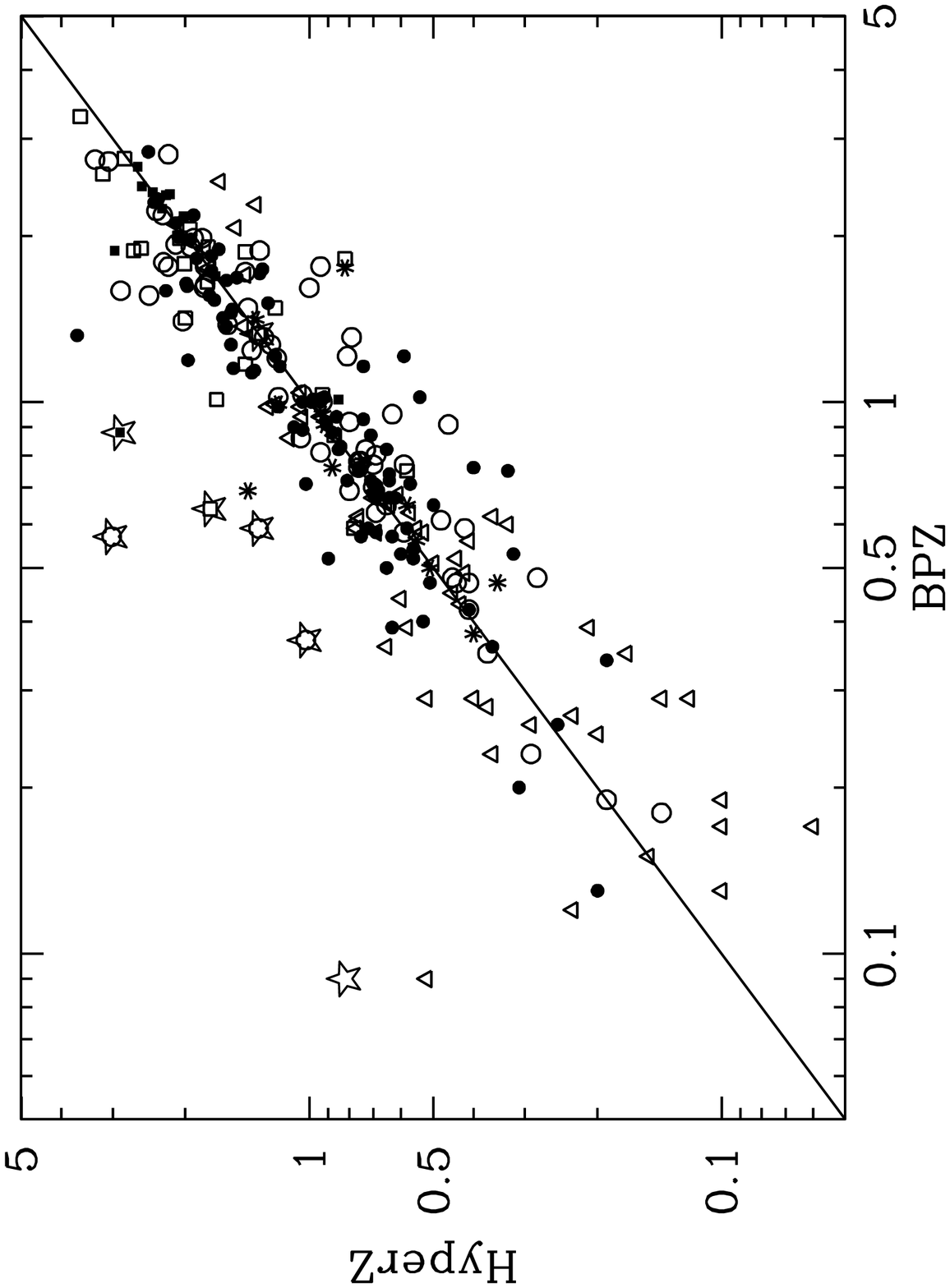}
\caption{~}
\end{figure}

\begin{figure}
\plotone{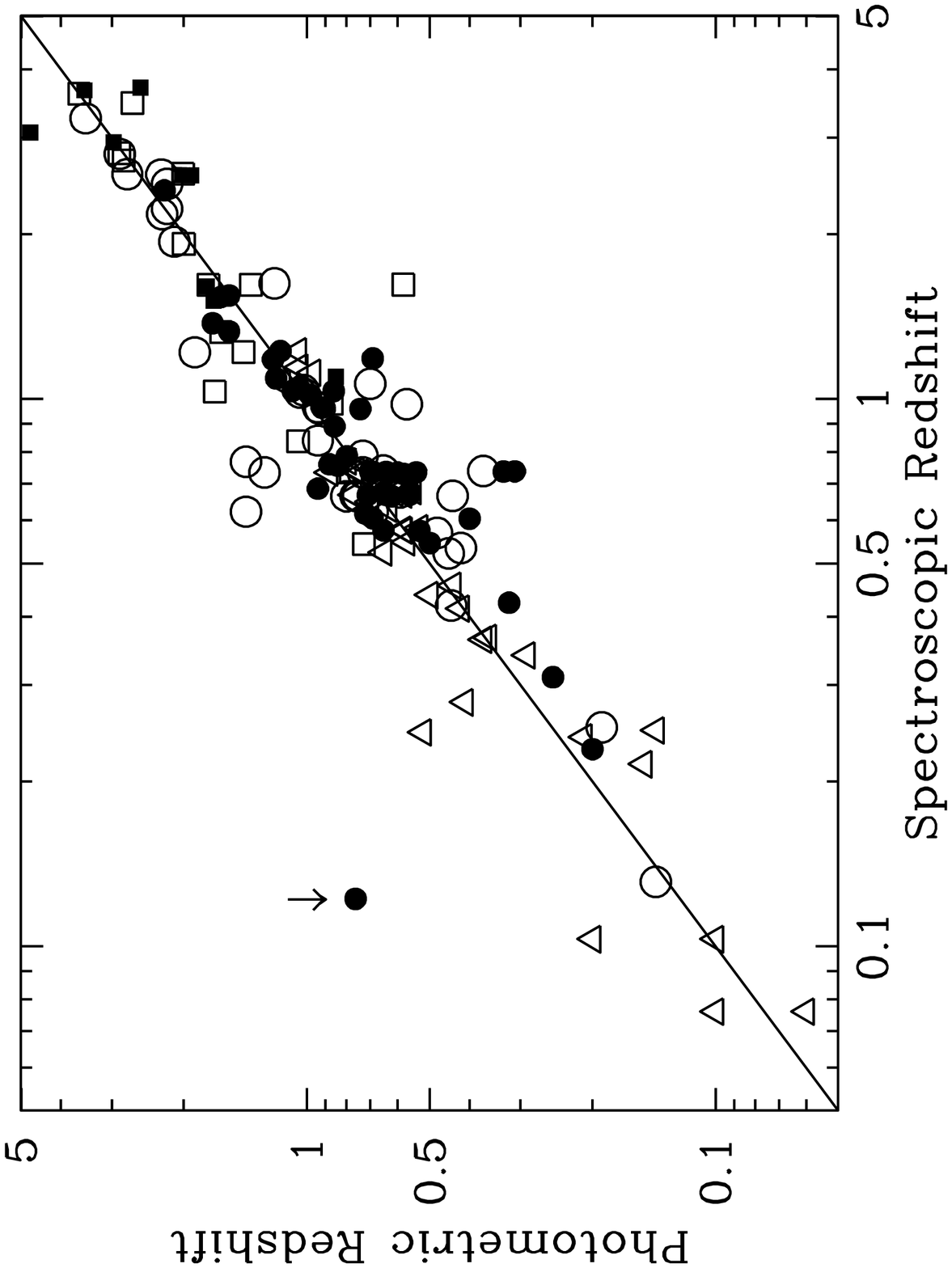}
\caption{~}
\end{figure}

\begin{figure}
\plotone{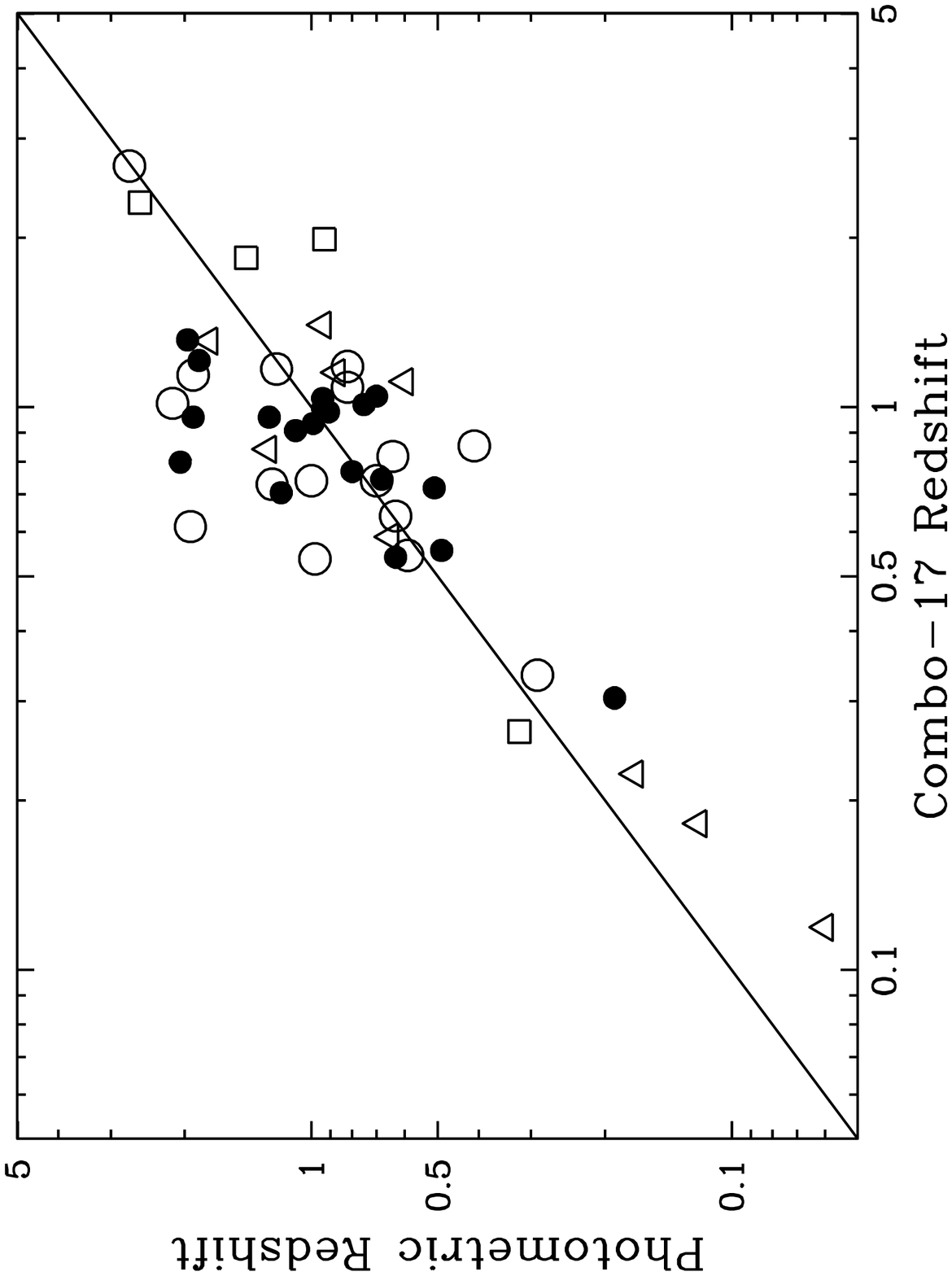}
\caption{~}
\end{figure}

\begin{figure}
\plotone{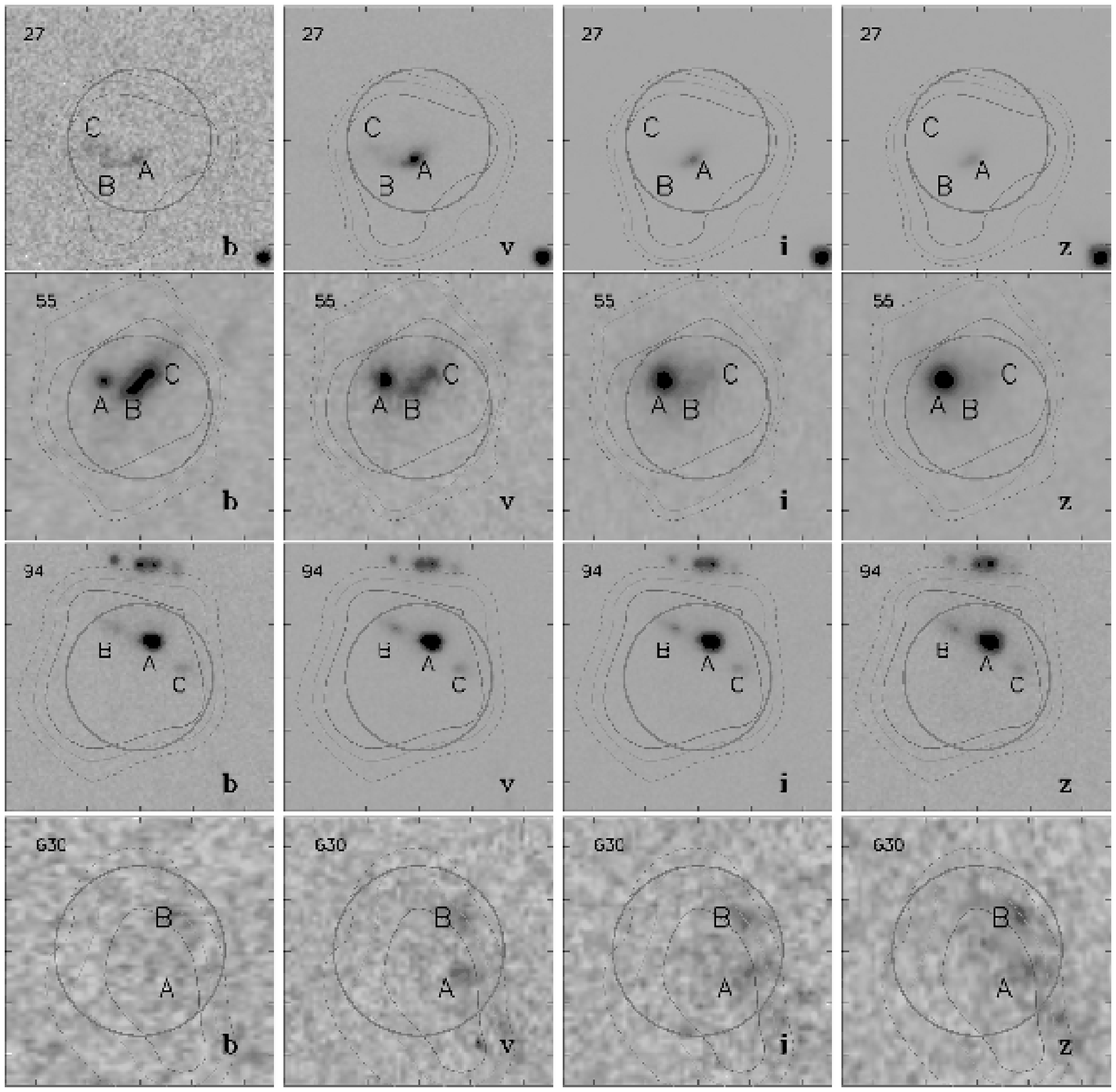}
\caption{~}
\end{figure}

\begin{figure}
\plotone{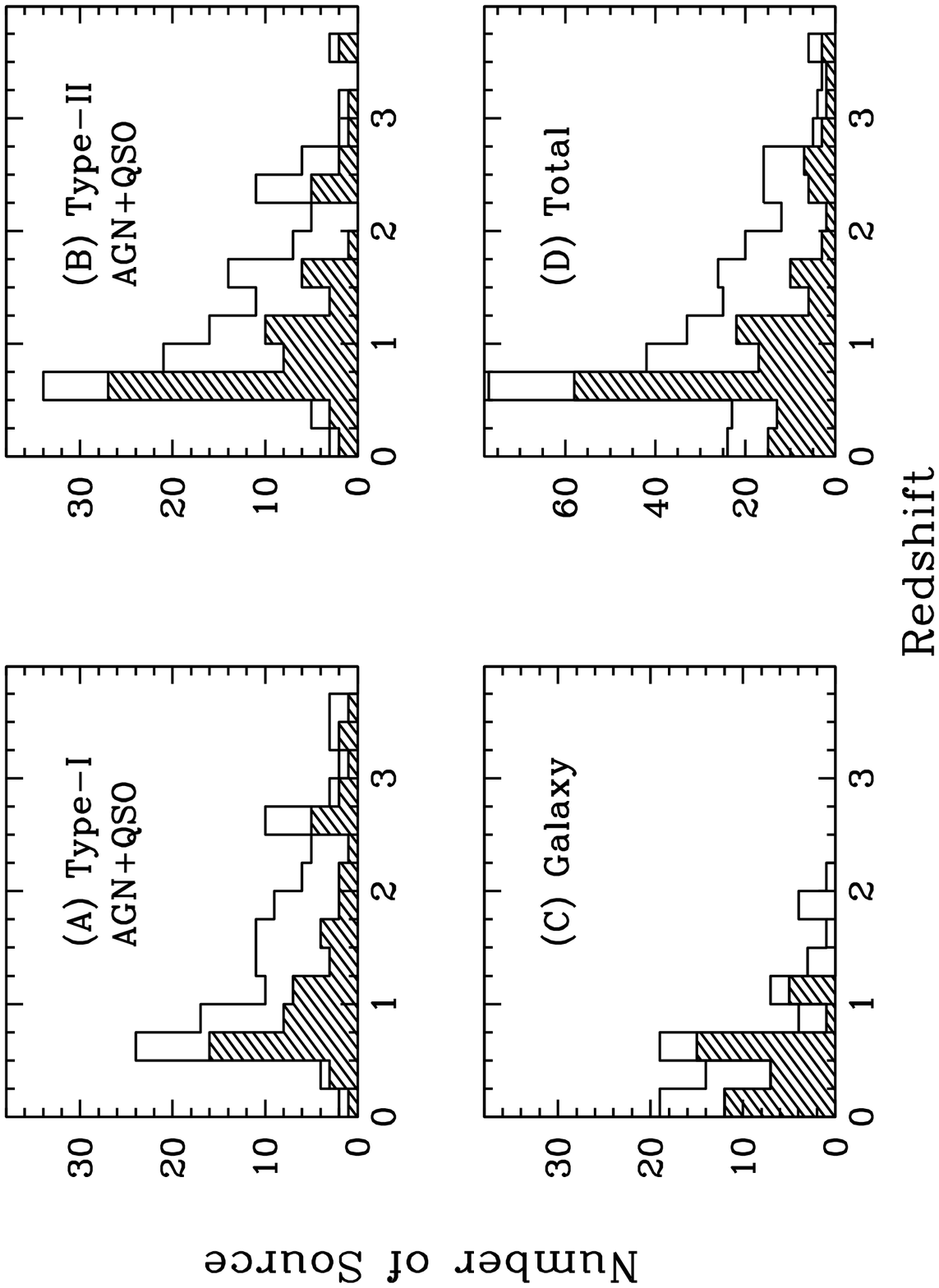}
\caption{~}
\end{figure}

\begin{figure}
\plotone{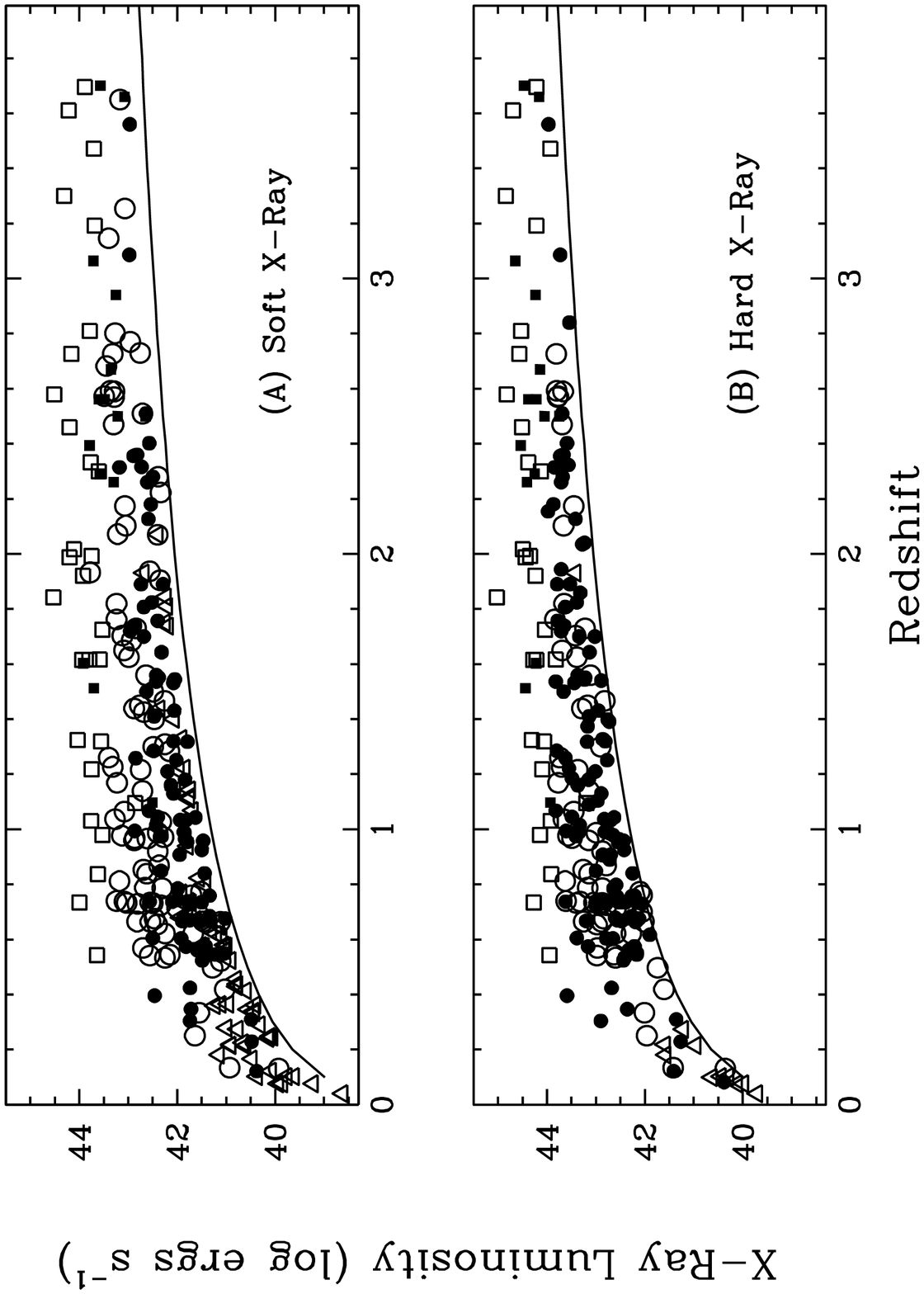}
\caption{~}
\end{figure}

\begin{figure}
\plotone{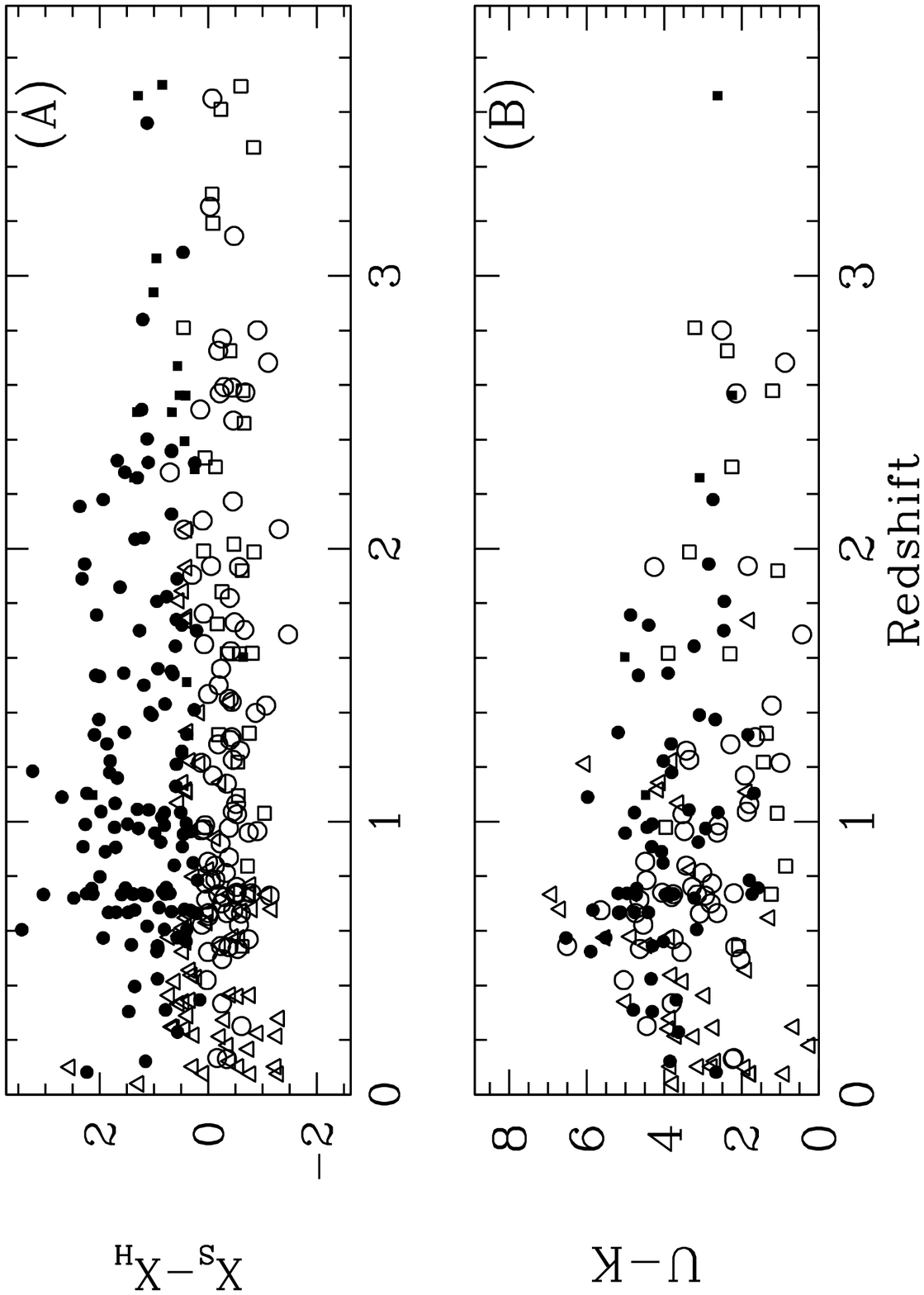}
\caption{~}
\end{figure}

\begin{figure}
\plotone{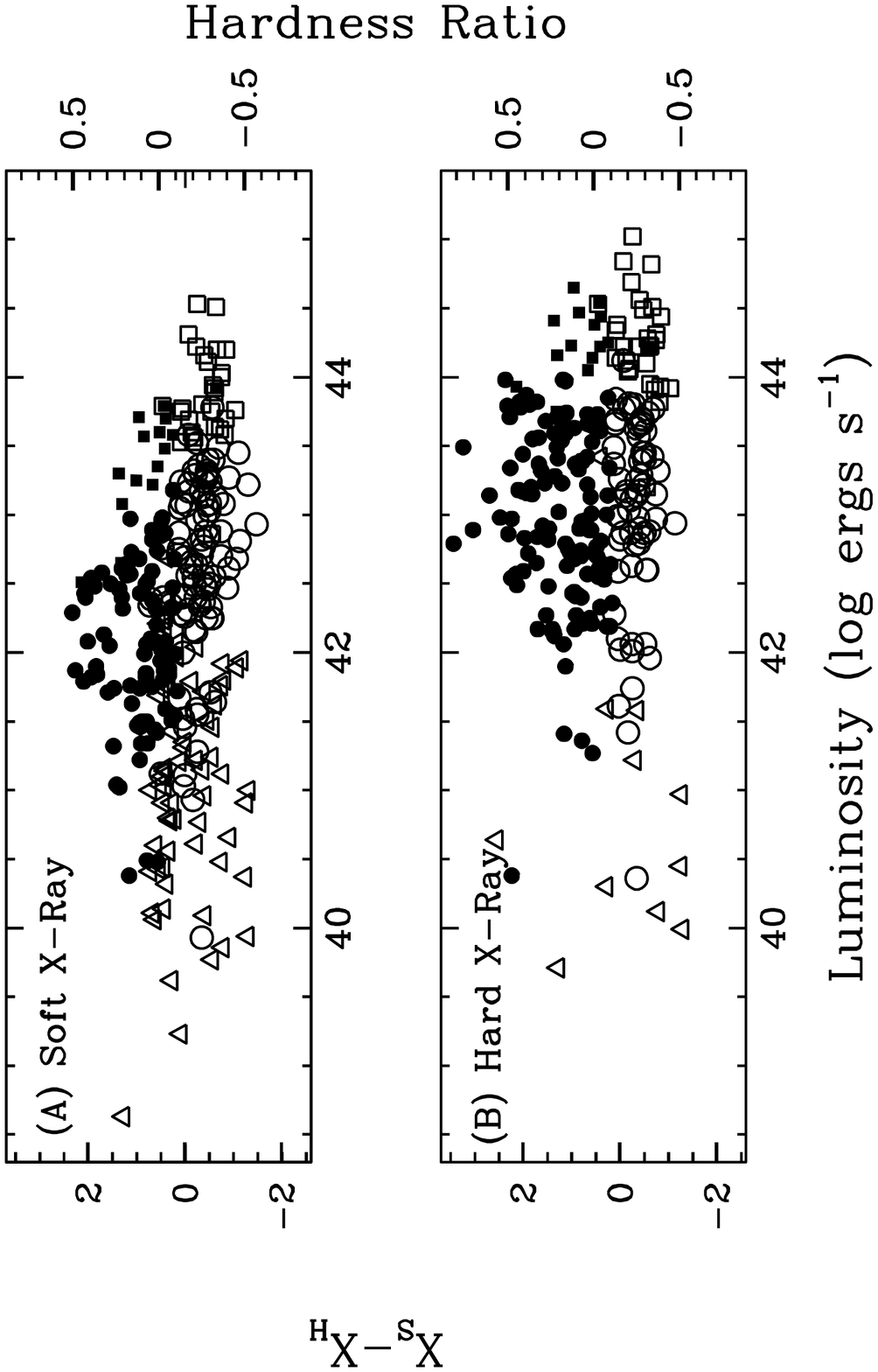}
\caption{~}
\end{figure}

\clearpage
{
\begin{deluxetable}{ccccccccc}
\tablecaption{Photometric Redshifts of X-Ray Sources in
Chandra Deep Field South\label{tbl-1}}
\tablewidth{0pt}
\footnotesize
\tighttable
\tablehead{
\colhead{X-ray ID} & 
\colhead{$\alpha$(2000)\tablenotemark{a}} & 
\colhead{$\delta$(J2000)\tablenotemark{a}}&
\colhead{Offset (\arcsec)} &
\colhead{Redshift\tablenotemark{b}}& \colhead{Range\tablenotemark{c}} & 
\colhead{Type}&
\colhead{Quality\tablenotemark{d}} &
\colhead{~}\\
}
\startdata
  1 &  03 33 09.56 & -27 46 03.9  &  0.9 &  0.35 &     \nodata   &AGN-2  &  1.6 & \\ 
  2 &  03 33 08.78 & -27 42 54.6  &  1.3 &  0.73 &  0.70 -  0.75 &AGN-1  &  0.4 & \\ 
  3 &  03 33 05.85 & -27 46 53.6  &  3.1 &  0.22 &  0.12 -  0.32 &Galaxy  &  0.4 & \\ 
  4 &  03 33 03.53 & -27 45 16.5  &  3.8 &  1.26 &     \nodata   &AGN-1  &  1.6 & \\ 
  6 &  03 33 02.68 & -27 48 23.4  &  0.8 &  2.46 &  2.39 -  2.50 &QSO-1  &  0.2 & \\ 
  7 &  03 33 01.71 & -27 58 18.9  &  1.0 &  1.84 &  0.53 -  1.95 &QSO-1  &  0.6\tablenotemark{g} & \\ 
  8 &  03 33 01.49 & -27 41 42.4  &  0.8 &  0.99 &  0.92 -  1.03 &AGN-2  &  0.9 & \\ 
  9 &  03 33 00.76 & -27 55 20.8  &  2.3 &  1.99 &  1.97 -  2.00 &QSO-1  &  0.4 & \\ 
 10 &  03 32 59.79 & -27 46 26.6  &  1.6 &  0.42 &     \nodata   &AGN-2  &  3.0 & \\ 
 11 &  03 32 59.85 & -27 47 48.4  &  1.5 &  2.58 &     \nodata   &QSO-1  &  3.0 & \\ 
 12 &  03 32 59.68 & -27 50 30.6  &  1.8 &  0.25 &     \nodata   &AGN-1  &  3.0 & \\ 
 13 &  03 32 59.07 & -27 43 39.8  &  1.2 &  0.73 &     \nodata   &AGN-1  &  3.0 & \\ 
 15 &  03 32 52.89 & -27 51 20.1  &  1.5 &  1.23 &     \nodata   &AGN-1  &  1.5 & \\ 
 17 &  03 32 49.11 & -27 55 06.7  &  2.6\tc  &  0.87 &  0.32 -  1.57 &AGN-1  &  0.6 & \\ 
 18 &  03 32 47.90 & -27 42 33.1  &  1.2 &  0.98 &     \nodata   &QSO-1  &  3.0 & \\ 
 19 &  03 32 47.92 & -27 41 48.2  &  1.0 &  0.74 &     \nodata   &AGN-1  &  3.0 & \\ 
 20 &  03 32 44.46 & -27 49 40.5  &  1.3 &  1.02 &     \nodata   &AGN-2  &  3.0 & \\ 
 21 &  03 32 44.32 & -27 52 51.5  &  1.5 &  3.47 &     \nodata   &QSO-1  &  3.0 & \\ 
 22 &  03 32 43.25 & -27 49 14.4  &  1.2 &  1.92 &     \nodata   &QSO-1  &  3.0 & \\ 
 23 &  03 32 41.72 & -27 44 01.5  &  3.7 &  0.73 &  0.33 -  1.04 &AGN-1  &  0.5 & \\ 
 24 &  03 32 41.86 & -27 52 02.7  &  1.6 &  3.61 &     \nodata   &QSO-1  &  3.0 & \\ 
 25 &  03 32 40.64 & -27 55 48.2  &  4.4 &  2.26 &   1.89 - 2.58   &QSO-2  &  0.5 & \\ 
 26 &  03 32 39.71 & -27 46 11.3  &  1.5\tc  &  1.65 &  1.36 -  1.89 &AGN-1  &  0.5 & \\ 
 27 &  03 32 39.59 & -27 48 51.9  &  2.4\tc  &  3.06 &     \nodata   &QSO-2  &  3.0 & \\ 
 28 &  03 32 39.10 & -27 46 02.0  &  1.0 &  1.22 &     \nodata   &AGN-1  &  3.0 & \\ 
 29 &  03 32 38.95 & -27 57 00.6  &  1.7 &  0.30 &  0.29 -  0.31 &AGN-2  &  0.9 & \\ 
 30 &  03 32 38.14 & -27 39 45.0  &  1.4 &  0.84 &     \nodata   &QSO-1  &  3.0 & \\ 
 31 &  03 32 37.78 & -27 52 12.6  &  1.3\tc  &  1.60 &     \nodata   &QSO-2  &  3.0 & \\ 
 32 &  03 32 37.47 & -27 40 00.3  &  1.0 &  0.66 &     \nodata   &AGN-1  &  3.0 & \\ 
 33 &  03 32 36.73 & -27 44 06.7  &  1.0 &  0.67 &     \nodata   &AGN-1  &  3.0 & \\ 
 34 &  03 32 34.96 & -27 55 11.2  &  1.8 &  0.84 &     \nodata   &AGN-1  &  3.0 & \\ 
 35 &  03 32 34.44 & -27 39 13.4  &  1.2 &  1.51 &     \nodata   &QSO-2  &  3.0 & \\ 
 36 &  03 32 32.98 & -27 45 45.9  &  2.9\tc  &  0.33 &  0.32 -  0.35 &AGN-1  &  0.9 & \\ 
 37 &  03 32 32.11 & -27 41 55.4  &  1.7\tc  &  0.99 &  0.62 -  1.06 &AGN-1  &  0.9 & \\ 
 38 &  03 32 30.23 & -27 45 04.8  &  1.2 &  0.74 &     \nodata   &AGN-1  &  3.0 & \\ 
 39 &  03 32 29.99 & -27 45 30.1  &  1.2 &  1.22 &     \nodata   &QSO-1  &  3.0 & \\ 
 40 &  03 32 29.01 & -27 57 30.5  &  1.5 &  0.55 &  0.50 -  0.58 &AGN-1  &  0.9 & \\ 
 41 &  03 32 27.64 & -27 41 45.1  &  0.9 &  0.67 &     \nodata   &AGN-2  &  3.0 & \\ 
 42 &  03 32 27.02 & -27 41 05.2  &  0.9 &  0.73 &     \nodata   &QSO-1  &  3.0 & \\ 
 43 &  03 32 26.78 & -27 41 45.8  &  1.6 &  0.74 &     \nodata   &AGN-2  &  3.0 & \\ 
 44 &  03 32 26.51 & -27 40 35.8  &  1.0 &  1.03 &     \nodata   &QSO-1  &  3.0 & \\ 
 45 &  03 32 25.71 & -27 43 05.8  &  0.8 &  2.29 &     2.14 - 2.60   &QSO-2  &  1.5 & \\ 
 46 &  03 32 25.17 & -27 42 19.0  &  1.4 &  1.62 &     \nodata   &QSO-1  &  3.0 & \\ 
 47 &  03 32 24.99 & -27 41 01.8  &  1.4 &  0.73 &     \nodata   &AGN-2  &  3.0 & \\ 
 48 &  03 32 24.87 & -27 56 00.1  &  1.7\tc  &  1.26 &  1.03 -  1.49 &AGN-2  &  0.5 & \\ 
 49 &  03 32 24.26 & -27 41 26.7  &  1.1 &  0.53 &     \nodata   &AGN-1  &  3.0 & \\ 
 50 &  03 32 19.03 & -27 47 55.4  &  1.1 &  0.67 &     \nodata   &AGN-2  &  1.9 & \\ 
 51 &  03 32 17.19 & -27 52 20.8  &  1.6\tc  &  1.10 &     \nodata   &QSO-2  &  3.0 & \\ 
 52 &  03 32 17.15 & -27 43 03.5  &  1.2 &  0.57 &     \nodata   &AGN-1  &  3.0 & \\ 
 53 &  03 32 14.99 & -27 51 27.1  &  1.7 &  0.68 &     \nodata   &AGN-1  &  3.0 & \\ 
 54 &  03 32 14.60 & -27 54 20.9  &  2.3 &  2.56 &     \nodata   &QSO-2  &  3.0 & \\ 
 55 &  03 32 14.02 & -27 51 00.9  &  1.4\tc  &  0.12 &     \nodata   &AGN-2  &  3.0 & \\ 
 56 &  03 32 13.26 & -27 42 41.1  &  0.7 &  0.60 &     \nodata   &AGN-2  &  3.0 & \\ 
 57 &  03 32 12.96 & -27 52 37.0  &  1.6 &  2.56 &     \nodata   &QSO-2  &  3.0 & \\ 
 58 &  03 32 11.80 & -27 46 28.4  &  1.0 &  0.92 &  0.58 -  1.22 &AGN-1  &  0.5 & \\ 
 59 &  03 32 11.37 & -27 52 13.9  &  1.9 &  0.97 &  0.32 -  4.11 &AGN-1  &  0.5 & \\ 
 60 &  03 32 10.93 & -27 44 15.1  &  1.0 &  1.61 &     \nodata   &QSO-1  &  3.0 & \\ 
 61 &  03 32 10.36 & -27 43 11.3  &  3.5 &  2.02 &  1.72 -  2.32 &QSO-1  &  0.5 & \\ 
 62 &  03 32 09.47 & -27 48 06.9  &  1.2 &  2.81 &     \nodata   &QSO-1  &  3.0 & \\ 
 63 &  03 32 08.68 & -27 47 34.5  &  1.2 &  0.54 &     \nodata   &QSO-1  &  3.0 & \\ 
 64 &  03 32 07.88 & -27 46 58.5  &  3.0 &  0.13 &  0.11 -  0.16 &AGN-1  &  0.4 & \\ 
 65 &  03 32 03.90 & -27 53 29.2  &  1.5 &  1.10 &  1.08 -  1.12 &QSO-1  &  0.5 & \\ 
 66 &  03 32 03.67 & -27 46 03.8  &  1.1 &  0.57 &     \nodata   &AGN-2  &  3.0 & \\ 
 67 &  03 32 02.47 & -27 46 00.5  &  0.9 &  1.62 &     \nodata   &QSO-1  &  3.0 & \\ 
 68 &  03 32 01.60 & -27 43 27.2  &  0.6 &  2.73 &     \nodata   &QSO-1  &  3.0 & \\ 
 69 &  03 32 01.47 & -27 41 38.9  &  0.5 &  0.85 &  0.57 -  1.14 &AGN-1  &  0.4 & \\ 
 70 &  03 32 01.23 & -27 46 47.3  &  3.8 &  1.07 &  0.88 -  1.15 &AGN-2  &  0.4 & \\ 
 71 &  03 32 00.37 & -27 43 19.9  &  1.1 &  1.04 &     \nodata   &AGN-1  &  3.0 & \\ 
 72 &  03 31 58.29 & -27 50 42.0  &  1.4 &  1.99 &  1.76 -  2.21 &QSO-1  &  0.5 & \\ 
 73 &  03 31 58.12 & -27 48 34.2  &  1.0 &  0.73 &     \nodata   &AGN-1  &  3.0 & \\ 
 74 &  03 31 57.80 & -27 42 08.9  &  0.8 &  0.65 &  0.59 -  0.72 &AGN-1  &  0.9 & \\ 
 75 &  03 31 55.39 & -27 54 48.3  &  0.8 &  0.74 &     \nodata   &AGN-2  &  3.0 & \\ 
 76 &  03 31 52.53 & -27 50 17.6  &  0.8 &  2.39 &     \nodata   &QSO-2  &  1.2 & \\ 
 77 &  03 33 01.53 & -27 45 42.7  &  2.2 &  0.62 &     \nodata   &AGN-1  &  3.0 & \\ 
 78 &  03 32 30.07 & -27 45 23.7  &  1.3\tc  &  0.96 &     \nodata   &AGN-1  &  3.0 & \\ 
 79 &  03 32 38.05 & -27 46 26.7  &  0.9 &  1.91 &  1.77 -  1.97 &AGN-1  &  0.5 & \\ 
 80 &  03 32 10.94 & -27 48 57.5  &  1.6 &  1.70 &  1.42 -  2.04 &AGN-1  &  0.5 & \\ 
 81 &  03 32 25.97 & -27 45 14.6  &  1.2 &  2.59 &  2.35 -  2.60 &AGN-1  &  0.5 & \\ 
 82 &  03 32 15.09 & -27 51 04.8  &  2.2 &  1.89 &  1.69 -  2.05 &AGN-2  &  0.5 & \\ 
 83 &  03 32 14.99 & -27 42 25.2  &  0.5\tc  &  1.76 &  1.68 -  1.85 &AGN-1  &  0.5 & \\ 
 84 &  03 32 46.77 & -27 42 12.2  &  1.7 &  0.10 &     \nodata   &Galaxy  &  3.0 & \\ 
 85 &  03 32 44.62 & -27 48 36.2  &  1.0\tc  &  2.59 &     \nodata   &AGN-1  &  1.2 & \\ 
 86 &  03 32 33.86 & -27 45 20.6  &  1.1 &  3.09 &  2.96 -  3.36 &AGN-2  &  0.5 & \\ 
 87 &  03 32 18.25 & -27 52 41.5  &  1.6 &  2.80 &     \nodata   &AGN-1  &  3.0 & \\ 
 89 &  03 32 08.28 & -27 41 53.8  &  0.5 &  2.47 &     \nodata   &AGN-1  &  3.0 & \\ 
 90 &  03 32 42.01 & -27 57 02.9  &  2.0 &\nodata&  \nodata &Star  &  3.0 & \\ 
 91 &  03 32 42.85 & -27 47 02.7  &  1.0\tc  &  3.19 &     \nodata   &QSO-1  &  1.0 & \\ 
 92 &  03 32 49.66 & -27 54 54.1  &  2.1 &\nodata&  \nodata &Star  &  3.0 & \\ 
 93 &  03 32 02.34 & -27 52 34.0  &  1.5 &  1.30 &  1.06 -  2.05 &AGN-1  &  0.5 & \\ 
 94 &  03 32 44.02 & -27 46 34.9  &  1.9\tc  &  1.69 &  1.56 -  1.89 &AGN-1  &  0.5 & \\ 
 95 &  03 32 29.89 & -27 44 24.4  &  0.8 &  0.08 &     \nodata   &Galaxy  &  3.0 & \\ 
 96 &  03 32 20.93 & -27 52 22.8  &  1.2 &  0.27 &  0.15 -  0.73 &Galaxy  &  0.6 & \\ 
 97 &  03 32 11.10 & -27 40 56.0  &  1.7 &  0.18 &     \nodata   &Galaxy  &  2.0 & \\ 
 98 &  03 32 44.28 & -27 51 41.4  &  1.0 &  0.28 &     \nodata   &Galaxy  &  3.0 & \\ 
 99 &  03 32 05.27 & -27 53 57.9  &  1.7 &  0.79 &  0.56 -  0.90 &AGN-1  &  0.5 & \\ 
100 &  03 32 36.00 & -27 48 50.6  &  0.9 &  1.31 &     \nodata   &AGN-1  &  1.9 & \\ 
101 &  03 32 55.50 & -27 47 53.5  &  1.9\tc  &  1.62 &     \nodata   &AGN-1  &  3.0 & \\ 
103 &  03 32 28.83 & -27 43 55.8  &  1.2 &  0.21 &     \nodata   &Galaxy  &  3.0 & \\ 
108 &  03 32 05.72 & -27 44 48.4  &  2.2 &  1.56 &  1.40 -  1.76 &AGN-1  &  0.5 & \\ 
110 &  03 32 58.61 & -27 46 32.2  &  2.0 &  0.62 &     \nodata   &AGN-1  &  3.0 & \\ 
112 &  03 31 51.96 & -27 53 27.3  &  1.8 &  2.94 &     \nodata   &QSO-2  &  3.0 & \\ 
114 &  03 32 07.63 & -27 52 13.8  &  1.4 &  1.72 &  1.55 -  1.92 &AGN-2  &  0.5 & \\ 
116 &  03 32 30.00 & -27 44 04.9  &  1.2 &  0.08 &     \nodata   &Galaxy  &  3.0 & \\ 
117 &  03 32 03.06 & -27 44 50.3  &  1.0 &  2.57 &     \nodata   &AGN-1  &  3.0 & \\ 
121 &  03 31 51.17 & -27 50 51.8  &  1.8 &  0.67 &     \nodata   &AGN-2  &  3.0 & \\ 
122 &  03 32 57.58 & -27 45 48.8  &  1.8 &  2.10 &  1.63 -  2.46 &AGN-1  &  0.5 & \\ 
124 &  03 32 02.45 & -27 45 25.6  &  3.0 &  0.61 &  0.29 -  0.93 &Galaxy  &  0.5 & \\ 
132 &  03 32 44.02 & -27 54 54.4  &  2.0\tc  &  0.91 &     \nodata   &AGN-2  &  1.4 & \\ 
133 &  03 32 02.79 & -27 44 28.8  &  3.5 &  1.21 &  0.35 -  6.95 &AGN-2  &  0.5 & \\ 
138 &  03 32 50.02 & -27 41 36.0  &  1.5 &  0.97 &     \nodata   &AGN-1  &  3.0 & \\ 
145 &  03 32 22.55 & -27 46 04.1  &  1.3 &  1.50 &  1.37 -  1.69 &AGN-2  &  0.5 & \\ 
146 &  03 32 47.05 & -27 53 33.4  &  2.4 &  2.67 &  2.47 -  2.85 &QSO-2  &  0.5 & \\ 
147 &  03 32 46.36 & -27 46 32.4  &  1.0 &  0.99 &  0.79 -  1.21 &AGN-2  &  0.5 & \\ 
148 &  03 32 35.21 & -27 53 18.0  &  1.8 &  1.74 &  1.50 -  2.02 &AGN-2  &  0.5 & \\ 
149 &  03 32 12.24 & -27 46 20.8  &  1.3\tc  &  1.03 &     \nodata   &AGN-2  &  3.0 & \\ 
150 &  03 32 25.17 & -27 54 50.2  &  1.7 &  1.09 &     \nodata   &AGN-2  &  3.0 & \\ 
151 &  03 32 20.49 & -27 47 32.5  &  1.3 &  0.60 &     \nodata   &AGN-2  &  3.0 & \\ 
152 &  03 32 59.33 & -27 48 59.1  &  1.5 &  1.28 &  1.20 -  2.26 &AGN-2  &  0.6 & \\ 
153 &  03 32 18.35 & -27 50 55.1  &  1.7 &  1.54 &     \nodata   &AGN-2  &  3.0 & \\ 
155 &  03 32 07.99 & -27 42 39.5  &  0.9 &  0.55 &     \nodata   &AGN-2  &  3.0 & \\ 
156 &  03 32 13.23 & -27 55 28.8  &  1.4 &  1.19 &     \nodata   &AGN-2  &  3.0 & \\ 
159 &  03 32 50.25 & -27 52 51.9  &  2.0\tc  &  3.30 &  3.04 - 3.62 &QSO-1  &  0.5 & \\ 
170 &  03 32 46.41 & -27 54 14.2  &  1.6 &  0.66 &     \nodata   &AGN-2  &  3.0 & \\ 
171 &  03 32 35.10 & -27 44 11.1  &  1.4 &  1.64 &     1.37 - 1.91   &AGN-2  &  0.5 & \\ 
173 &  03 32 16.75 & -27 43 27.7  &  1.3 &  0.52 &     \nodata   &Galaxy  &  3.0 & \\ 
174 &  03 33 01.20 & -27 44 20.9  &  0.9 &  1.55 &  1.45 -  1.61 &AGN-2  &  0.5 & \\ 
175 &  03 32 51.83 & -27 44 37.0  &  1.8 &  0.52 &     \nodata   &AGN-1  &  3.0 & \\ 
176 &  03 33 09.24 & -27 44 50.0  &  1.7 &  0.79 &     \nodata   &AGN-1  &  3.0 & \\ 
177 &  03 32 56.96 & -27 50 08.9  &  1.5 &  1.14 &     \nodata   &Galaxy  &  3.0 & \\ 
178 &  03 32 13.89 & -27 50 33.5  &  1.3 &  0.29 &  0.17 -  6.79 &Galaxy  &  0.5 & \\ 
179 &  03 31 49.51 & -27 50 34.2  &  1.3 &  2.73 &  2.56 -  2.90 &AGN-1  &  0.5 & \\ 
183 &  03 32 34.19 & -27 56 44.8  &  3.6 &  0.08 &  0.01 -  0.21 &AGN-2  &  0.5 & \\ 
184 &  03 32 48.19 & -27 52 56.9  &  1.7 &  0.67 &     \nodata   &AGN-2  &  3.0 & \\ 
185 &  03 32 10.93 & -27 43 43.2  &  1.5 &  0.93 &  0.83 -  1.05 &AGN-2  &  0.9 & \\ 
186 &  03 32 52.36 & -27 45 56.8  &  1.2\tc  &  1.11 &  0.92 -  1.30 &Galaxy  &  0.4 & \\ 
188 &  03 32 22.56 & -27 49 49.9  &  1.2 &  0.73 &     \nodata   &AGN-2  &  3.0 & \\ 
189 &  03 32 45.75 & -27 42 13.0  &  2.1 &  0.76 &     \nodata   &AGN-2  &  3.0 & \\ 
190 &  03 32 35.86 & -27 40 59.7  &  2.2 &  0.73 &     \nodata   &AGN-2  &  3.0 & \\ 
200 &  03 32 54.97 & -27 45 07.1  &  0.9 &  0.85 &  0.75 -  1.38 &AGN-2  &  0.4 & \\ 
201 &  03 32 39.06 & -27 44 39.3  &  1.4 &  0.68 &     \nodata   &AGN-2  &  3.0 & \\ 
202 &  03 32 29.87 & -27 51 06.0  &  1.3 &  3.70 &     \nodata   &QSO-2  &  3.0 & \\ 
203 &  03 32 26.69 & -27 40 13.6  &  0.6\tc  &  1.17 &  0.92 -  1.50 &AGN-1  &  0.7 & \\ 
204 &  03 32 23.17 & -27 45 54.9  &  0.9 &  1.22 &     \nodata   &Galaxy  &  3.0 & \\ 
205 &  03 32 17.11 & -27 41 37.2  &  1.3 &  1.56 &     1.31 - 2.30   &AGN-2  &  0.5 & \\ 
206 &  03 32 16.21 & -27 39 30.5  &  1.2 &  1.32 &     \nodata   &QSO-1  &  3.0 & \\ 
207 &  03 32 07.91 & -27 37 33.4  &  2.8 &  0.40 &  0.30 -  0.50 &AGN-2  &  0.4 & \\ 
208 &  03 31 52.54 & -27 46 42.5  &  0.9 &  0.72 &  0.63 -  0.81 &AGN-1  &  0.6 & \\ 
209 &  03 31 47.29 & -27 53 13.8  &  1.3 &  1.32 &  1.07 -  1.73 &QSO-1  &  0.5 & \\ 
210 &  03 32 38.33 & -27 55 53.3  &  1.8 &  1.73 &  1.47 -  2.17 &AGN-1  &  0.5 & \\ 
211 &  03 32 05.92 & -27 54 49.7  &  1.7 &  0.68 &     \nodata   &Galaxy  &  3.0 & \\ 
213 &  03 32 00.45 & -27 53 56.1  &  2.8 &  0.60 &  0.19 -  1.73 &AGN-2  &  0.5 & \\ 
217 &  03 32 33.03 & -27 52 02.9  &  4.4 &  3.65 &  3.21 -  4.38 &AGN-1  &  
0.2\tablenotemark{h} & \\ 
218 &  03 32 16.37 & -27 52 01.3  &  2.3\tc  &  0.50 &  \nodata   &AGN-1  &  1.4 & \\ 
219 &  03 31 50.42 & -27 51 52.1  &  1.7 &  1.73 &  1.55 -  1.92 &QSO-1  &  0.5 & \\ 
220 &  03 32 32.76 & -27 51 51.3  &  1.4 &  1.40 &  1.37 -  1.42 &AGN-1  &  0.5 & \\ 
221 &  03 32 08.83 & -27 44 23.8  &  2.5 &  2.51 &  2.17 -  2.73 &AGN-1  &  0.5 & \\ 
222 &  03 32 54.53 & -27 45 02.1  &  0.8 &  1.14 &  0.88 -  1.40 &AGN-1  &  0.4 & \\ 
224 &  03 32 28.77 & -27 46 20.7  &  0.8 &  0.74 &     \nodata   &Galaxy  &  3.0 & \\ 
225 &  03 31 49.42 & -27 46 34.4  &  1.3 &  2.30 &  2.27 -  2.33 &QSO-1  &  0.6 & \\ 
226 &  03 32 04.42 & -27 46 43.2  &  2.0 &  1.45 &  0.83 -  2.29 &AGN-1  &  0.5 & \\ 
227 &  03 32 05.47 & -27 46 46.8  &  2.2 &  2.18 &  1.78 -  2.54 &AGN-2  &  0.5 & \\ 
229 &  03 32 56.34 & -27 48 34.1  &  2.1 &  0.10 &     \nodata   &Galaxy  &  3.0 & \\ 
230 &  03 31 53.55 & -27 48 43.1  &  1.4 &  2.17 &     \nodata   &AGN-1  &  3.0 & \\ 
232 &  03 31 55.84 & -27 49 21.4  &  1.3 &  0.94 &  0.69 -  1.20 &Galaxy  &  0.6 & \\ 
233 &  03 32 25.76 & -27 49 36.4  &  1.7 &  0.58 &     \nodata   &Galaxy  &  3.0 & \\ 
236 &  03 32 11.46 & -27 50 06.7  &  1.1 &  0.76 &  0.70 -  1.26 &AGN-1  &  0.6 & \\ 
237 &  03 32 58.52 & -27 50 08.1  &  1.5 &\nodata&  \nodata &Star  &  3.0 & \\ 
238 &  03 31 47.98 & -27 50 45.4  &  1.5 &  1.06 &     \nodata   &AGN-1  &  3.0 & \\ 
239 &  03 32 36.18 & -27 51 26.8  &  1.4 &  1.47 &  1.22 -  1.72 &AGN-1  &  0.5 & \\ 
240 &  03 32 58.96 & -27 51 41.8  &  3.8\tc  &  1.41 &  1.29 -  1.65 &AGN-2  &  0.5 & \\ 
241 &  03 32 24.22 & -27 42 58.0  &  1.1 &  0.70 &  0.64 -  0.87 &AGN-1  &  0.5 & \\ 
242 &  03 32 51.86 & -27 42 29.7  &  1.1 &  1.03 &     \nodata   &AGN-1  &  3.0 & \\
243 &  03 32 08.41 & -27 40 47.1  &  1.3 &  2.50 &  1.50 -  3.50 &QSO-2  &  0.3\tablenotemark{i} & \\ 
244 &  03 32 04.33 & -27 40 26.7  &  1.1 &  0.97 &  0.31 -  3.95 &AGN-1  &  0.5 & \\
246 &  03 32 22.86 & -27 39 36.9  &  1.3 &  0.71 &  0.64 -  0.79 &AGN-1  &  0.5 & \\ 
247 &  03 32 35.09 & -27 55 33.2  &  3.2 &  0.04 &     \nodata   &Galaxy  &  3.0 & \\ 
248 &  03 32 10.24 & -27 54 16.4  &  2.1 &  0.69 &     \nodata   &AGN-2  &  3.0 & \\ 
249 &  03 32 19.29 & -27 54 06.2  &  3.0 &  0.96 &     \nodata   &AGN-2  &  2.0 & \\ 
251 &  03 32 07.27 & -27 52 29.2  &  1.1 &  2.13 &  2.03 -  2.29 &AGN-2  &  0.5 & \\ 
252 &  03 32 47.05 & -27 43 46.6  &  0.7\tc  &  1.18 &     \nodata   &AGN-2  &  3.0 & \\ 
253 &  03 32 20.05 & -27 44 47.7  &  1.3\tc  &  1.89 &     \nodata   &AGN-2  &  1.9 & \\ 
254 &  03 32 19.88 & -27 45 18.2  &  1.5\tc  &  0.10 &  0.01 -  0.12 &Galaxy  &  0.7 & \\ 
256 &  03 32 43.07 & -27 48 45.1  &  1.0 &  1.53 &  1.47 -  1.58 &AGN-2  &  0.5 & \\ 
257 &  03 32 13.39 & -27 48 57.8  &  1.6 &  0.55 &     \nodata   &AGN-2  &  1.5 & \\ 
259 &  03 32 06.14 & -27 49 27.8  &  1.4 &  1.76 &  1.63 -  1.88 &AGN-2  &  0.5 & \\ 
260 &  03 32 25.12 & -27 50 43.2  &  1.6 &  1.04 &     \nodata   &AGN-2  &  3.0 & \\ 
263 &  03 32 18.87 & -27 51 34.4  &  2.4 &  3.66 &     \nodata   &QSO-2  &  3.0 & \\ 
264 &  03 32 29.76 & -27 51 46.9  &  0.6 &  1.32 &     \nodata   &AGN-2  &  1.9 & \\ 
265 &  03 32 33.31 & -27 42 36.4  &  1.2\tc  &  1.16 &  1.02 - 1.32 &AGN-2  &  0.5 & \\ 
266 &  03 32 13.85 & -27 42 49.1  &  1.7 &  0.73 &     \nodata   &AGN-2  &  3.0 & \\ 
267 &  03 32 04.85 & -27 41 27.5  &  1.7\tc  &  0.72 &     \nodata   &AGN-2  &  1.2 & \\ 
268 &  03 32 49.22 & -27 40 50.6  &  1.8 &  1.22 &     \nodata   &AGN-2  &  3.0 & \\ 
501 &  03 33 10.19 & -27 48 42.2  &  2.0 &  0.81 &  0.75 -  0.87 &AGN-1  &  0.6 & \\ 
502 &  03 33 08.17 & -27 50 33.4  &  1.4 &  0.73 &  0.65 -  0.81 &AGN-2  &  0.6 & \\ 
503 &  03 33 07.62 & -27 51 27.2  &  1.9 &  0.54 &  0.34 -  0.74 &AGN-1  &  0.4 & \\ 
504 &  03 33 05.67 & -27 52 14.5  &  2.0 &  0.52 &  0.00 -  0.66 &AGN-2  &  0.6 & \\ 
505 &  03 33 04.80 & -27 47 31.9  &  2.5 &  2.26 &  2.06 -  2.44 &AGN-2  &  0.5 & \\ 
506 &  03 33 03.04 & -27 51 45.8  &  2.1 &  3.69 &  3.12 -  4.19 &QSO-1  &  0.5 & \\ 
507 &  03 33 00.14 & -27 49 23.4  &  2.1 &  0.99 &  0.88 -  1.00 &AGN-2  &  0.6 & \\ 
508 &  03 32 51.66 & -27 52 13.0  &  2.2 &  2.50 &  1.50 -  3.50 &QSO-2  &  0.5 & \\
509 &  03 32 42.22 & -27 57 51.6  &  3.4 &  0.56 &  0.46 -  0.60 &AGN-2  &  0.6 & \\ 
510 &  03 32 38.79 & -27 51 22.1  &  0.9\tc  &  2.51 &  2.17 -  2.67 &AGN-2  &  0.5 & \\ 
511 &  03 32 36.44 & -27 46 31.6  &  2.9 &  0.77 &     \nodata   &AGN-1  &  2.0 & \\ 
512 &  03 32 34.36 & -27 43 50.3  &  1.5 &  0.67 &     \nodata   &AGN-2  &  3.0 & \\ 
513 &  03 32 34.00 & -27 48 59.7  &  1.3 &  3.52 &  3.26 -  3.80 &AGN-2  &  0.5 & \\ 
514 &  03 32 33.47 & -27 43 12.9  &  1.5 &  0.10 &     \nodata   &Galaxy  &  3.0 & \\ 
515 &  03 32 32.21 & -27 46 51.7  &  0.8 &  2.19 &  2.15 -  2.45 &AGN-2  &  0.5 & \\ 
516 &  03 32 31.37 & -27 47 25.1  &  2.8 &  0.67 &     \nodata   &AGN-1  &  3.0 & \\ 
517 &  03 32 30.14 & -28 00 24.4  &  2.5 &  2.33 &  0.00 -  2.88 &QSO-1  &  0.6 & \\ 
518 &  03 32 26.78 & -27 46 04.0  &  1.5 &  0.84 &  0.32 -  1.90 &AGN-2  &  0.5 & \\ 
519 &  03 32 25.87 & -27 55 06.5  &  2.4\tc  &  1.03 &     \nodata   &AGN-2  &  3.0 & \\ 
520 &  03 32 25.94 & -27 39 27.7  &  0.3\tc  &  0.79 &     \nodata   &AGN-2  &  3.0 & \\ 
521 &  03 32 22.78 & -27 52 24.0  &  1.4 &  0.13 &     \nodata   &AGN-1  &  3.0 & \\ 
522 &  03 32 21.42 & -27 55 49.5  &  1.9\tc  &  2.57 &     \nodata   &AGN-1  &  2.0 & \\ 
523 &  03 32 20.49 & -27 42 27.2  &  1.9 &  1.32 &  0.32 -  8.5 &AGN-2  &  0.3\tablenotemark{j} & \\ 
524 &  03 32 19.95 & -27 42 43.1  &  1.5\tc  &  2.36 &  1.98 -  2.80 &AGN-2  &  0.5 & \\ 
525 &  03 32 19.81 & -27 41 22.9  &  1.3 &  0.23 &     \nodata   &AGN-2  &  3.0 & \\ 
526 &  03 32 18.71 & -27 44 12.8  &  1.7\tc  &  0.96 &     \nodata   &AGN-2  &  3.0 & \\ 
527 &  03 32 18.37 & -27 54 12.1  &  4.0 &  4.49 &  3.41 -  5.14 &AGN-1 &  0.5 & \\ 
528 &  03 32 17.14 & -27 54 02.6  &  1.2\tc  &  1.43 &  0.82 -  1.96 &AGN-2  &  0.5 & \\ 
529 &  03 32 16.41 & -27 55 24.4  &  2.0 &  0.73 &  0.58 -  0.89 & AGN-2  &  0.6 & \\ 
530 &  03 32 14.91 & -27 38 43.8  &  0.2 &  1.04 &  0.88 -  1.21 &AGN-2  &  0.6 & \\ 
531 &  03 32 14.44 & -27 51 10.9  &  1.4 &  1.54 &     \nodata   &AGN-2  &  3.0 & \\ 
532 &  03 32 14.08 & -27 42 30.2  &  2.0\tc  &  0.95 &  0.54 -  1.37 &AGN-2  &  0.9 & \\ 
533 &  03 32 14.03 & -27 56 01.6  &  1.1 &  0.54 &  0.43 -  0.77 &AGN-2  &  0.5 & \\ 
534 &  03 32 12.21 & -27 45 30.3  &  0.6 &  0.68 &     \nodata   &AGN-2  &  3.0 & \\ 
535 &  03 32 11.42 & -27 46 50.2  &  1.1\tc  &  0.57 &     \nodata   &AGN-2  &  3.0 & \\ 
536 &  03 32 10.77 & -27 42 34.8  &  1.9\tc  &  0.42 &     \nodata   &AGN-1  &  3.0 & \\ 
537 &  03 32 09.83 & -27 50 14.1  &  2.6 &  1.54 &  1.08 -  1.72 &AGN-2  &  0.5 & \\ 
538 &  03 32 08.54 & -27 46 48.4  &  1.4 &  0.31 &     \nodata   &AGN-2  &  3.0 & \\ 
539 &  03 32 04.07 & -27 37 25.7  &  0.6 &  0.98 &     \nodata   &AGN-1  &  3.0 & \\ 
540 &  03 32 02.57 & -27 50 52.8  &  1.4 &  1.25 &  1.16 -  1.34 &AGN-2  &  0.5 & \\ 
541 &  03 31 59.59 & -27 49 47.8  &  2.7 &  1.82 &  1.13 -  2.68 &AGN-2  &  0.5 & \\ 
542 &  03 31 58.43 & -27 54 35.7  &  1.6 &  1.70 &  1.04 -  3.15 &AGN-2  &  0.5 & \\ 
543 &  03 31 56.99 & -27 51 00.6  &  1.5 &  1.81 &  1.64 -  1.90 &AGN-2  &  0.5 & \\ 
544 &  03 31 54.48 & -27 51 05.7  &  2.6 &  2.36 &  2.14 -  2.57 &AGN-2  &  0.5 & \\ 
545 &  03 31 54.41 & -27 41 59.0  &  1.5 &  0.97 &  0.93 -  2.47 &AGN-2  &  0.6 & \\ 
546 &  03 31 52.33 & -27 47 52.5  &  1.5 &  2.31 &  2.19 -  2.38 &AGN-2  &  0.5 & \\ 
547 &  03 31 50.41 & -27 52 37.9  &  2.4 &  2.32 &     \nodata   &AGN-2  &  1.2 & \\ 
548 &  03 31 44.88 & -27 51 59.1  &  2.5 &  1.44 &  1.41 -  1.46 &AGN-1  &  0.5 & \\ 
549 &  03 32 22.57 & -27 58 05.7  &  2.1 &\nodata&  \nodata &Star  &  3.0 & \\ 
550 &  03 33 00.60 & -27 57 47.6  &  3.4 &  1.93 &  1.81 -  2.06 &AGN-1  &  0.5 & \\ 
551 &  03 32 16.15 & -27 56 44.3  &  1.8 &  2.68 &  2.57 -  2.84 &AGN-1  &  0.6 & \\ 
552 &  03 32 15.81 & -27 53 24.8  &  1.3 &  0.67 &     \nodata   &Galaxy  &  3.0 & \\ 
553 &  03 32 56.66 & -27 53 16.7  &  2.5 &  0.37 &     \nodata   &Galaxy  &  3.0 & \\ 
554 &  03 31 50.77 & -27 53 01.1  &  1.5 &  0.23 &  0.05 -  3.13 &Galaxy  &  0.6 & \\ 
555 &  03 32 37.97 & -27 53 08.0  &  1.6 &  2.28 &  1.94 -  2.64 &AGN-1  &  0.5 & \\ 
556 &  03 32 00.44 & -27 52 28.8  &  0.8 &  0.63 &     \nodata   &Galaxy  &  3.0 & \\ 
557 &  03 32 37.99 & -27 43 59.4  &  2.7 &  1.81 &  1.69 -  2.11 &Galaxy  &  0.5 & \\ 
558 &  03 31 58.16 & -27 44 59.6  &  0.8 &  0.57 &     \nodata   &Galaxy  &  3.0 & \\ 
559 &  03 32 57.13 & -27 45 34.4  &  2.0 &  0.01 &  0.00 -  0.02 &Galaxy  &  0.6 & \\ 
560 &  03 32 06.28 & -27 45 36.9  &  1.1 &  0.67 &     \nodata   &Galaxy  &  3.0 & \\ 
561 &  03 32 22.44 & -27 45 43.7  &  1.1\tc  &  0.62 &  0.38 - 1.12 &Galaxy  &  0.5 & \\ 
562 &  03 31 51.51 & -27 45 54.1  &  0.6 &  0.36 &  0.06 -  1.15 &Galaxy  &  0.5 & \\ 
563 &  03 32 31.47 & -27 46 23.2  &  1.7\tc  &  2.22 &     \nodata   &AGN-1  &  3.0 & \\ 
564 &  03 32 16.71 & -27 46 38.1  &  1.5 &  0.43 &  0.03 -  0.65 &Galaxy  &  0.5 & \\ 
565 &  03 32 24.87 & -27 47 06.4  &  1.3 &  0.36 &     \nodata   &Galaxy  &  3.0 & \\ 
566 &  03 32 18.03 & -27 47 18.7  &  1.4\tc  &  0.73 &     \nodata   &Galaxy  &  2.0 & \\ 
567 &  03 32 38.80 & -27 47 32.5  &  1.2 &  0.46 &     \nodata   &Galaxy  &  3.0 & \\ 
568 &  03 33 10.91 & -27 47 36.6  &  4.0 &  3.15 &  2.78 -  3.37 &AGN-1  &  0.5 & \\ 
569 &  03 31 48.06 & -27 48 01.9  &  1.3 &  2.07 &  1.96 -  2.19 &AGN-1  &  0.5 & \\ 
570 &  03 32 22.55 & -27 48 04.5  &  1.5 &  1.28 &  1.04 -  1.76 &AGN-1  &  0.5 & \\ 
571 &  03 33 03.77 & -27 48 10.9  &  1.2 &  1.44 &  1.42 -  1.45 &Galaxy  &  0.5 & \\ 
572 &  03 32 22.19 & -27 48 11.4  &  1.2 &  2.73 &  1.69 -  2.63 &AGN-1  &  0.5 & \\ 
573 &  03 32 44.45 & -27 48 19.3  &  1.0 &  0.41 &     \nodata   &Galaxy  &  3.0 & \\ 
574 &  03 32 31.56 & -27 48 53.8  &  1.9 &  1.84 &  1.65 -  2.05 &Galaxy  &  0.5 & \\ 
575 &  03 32 17.08 & -27 49 21.9  &  1.6 &  0.34 &     \nodata   &Galaxy  &  3.0 & \\ 
576 &  03 31 44.00 & -27 49 28.3  &  4.4 &  1.50 &  1.37 -  1.66 &AGN-1  &  0.5 & \\ 
577 &  03 32 36.19 & -27 49 32.0  &  1.8\tc  &  0.55 &     \nodata   &Galaxy  &  3.0 & \\ 
578 &  03 32 48.55 & -27 49 34.9  &  0.7 &  1.12 &     \nodata   &Galaxy  &  3.0 & \\ 
579 &  03 32 34.06 & -27 49 37.8  &  2.1\tc  &  0.82 &  0.54 -  1.00 &Galaxy  &  0.9 & \\
580 &  03 32 15.99 & -27 49 43.4  &  2.7\tc  &  0.66 &     \nodata   &AGN-1  &  3.0 & \\ 
581 &  03 32 07.37 & -27 49 42.0  &  2.2\tc  &  0.80 &  0.17 -  0.90 &Galaxy &  0.6 & \\ 
582 &  03 32 38.83 & -27 49 56.5  &  1.0 &  0.24 &     \nodata   &Galaxy  &  3.0 & \\ 
583 &  03 32 13.86 & -27 50 00.5  &  2.1 &  2.77 &  2.55 -  2.99 &AGN-1  &  0.5 & \\ 
584 &  03 32 17.86 & -27 50 07.0  &  1.4 &\nodata&  \nodata &Star  &  3.0 & \\ 
585 &  03 31 55.53 & -27 50 29.7  &  2.9 &  1.21 &     \nodata   &Galaxy  &  1.6 & \\ 
586 &  03 32 39.48 & -27 50 32.0  &  1.1 &  0.58 &     \nodata   &Galaxy  &  3.0 & \\ 
587 &  03 32 15.29 & -27 50 39.4  &  2.1 &  0.25 &     \nodata   &Galaxy  &  3.0 & \\ 
588 &  03 31 55.62 & -27 50 44.0  &  0.4 &\nodata&  \nodata &Star  &  3.0 & \\ 
589 &  03 32 25.83 & -27 51 20.3  &  1.3 &  1.33 &  0.45 -  8.50 &Galaxy  &  0.5 & \\ 
590 &  03 32 07.12 & -27 51 28.7  &  2.4 &  0.35 &  0.05 -  1.08 &Galaxy  &  0.5 & \\ 
591 &  03 31 44.75 & -27 51 37.1  &  4.1 &  1.43 &  1.33 -  1.47 &AGN-1  &  0.5 & \\ 
592 &  03 32 47.19 & -27 51 47.8  &  1.3 &  1.07 &     \nodata   &Galaxy  &  3.0 & \\ 
593 &  03 32 14.79 & -27 44 02.5  &  1.4 &  2.07 &  1.59 -  2.23 &Galaxy  &  0.5 & \\ 
594 &  03 32 09.71 & -27 42 48.4  &  2.2 &  0.73 &     \nodata   &Galaxy  &  2.0 & \\ 
595 &  03 32 15.77 & -27 39 54.1  &  1.0 &  0.36 &  0.19 -  0.98 &Galaxy  &  0.5 & \\ 
596 &  03 32 31.81 & -27 57 14.3  &  1.7 &  1.94 &  1.84 -  1.99 &AGN-2  &  0.5 & \\ 
597 &  03 32 51.35 & -27 55 44.0  &  0.6 &  2.32 &  2.24 -  2.40 &AGN-2  &  0.5 & \\ 
598 &  03 32 24.69 & -27 54 11.6  &  1.7 &  0.62 &     \nodata   &AGN-2  &  3.0 & \\ 
599 &  03 32 29.80 & -27 53 28.6  &  1.8 &  2.84 &  2.12 -  3.00 &AGN-2  &  0.5 & \\ 
600 &  03 32 13.84 & -27 45 25.9  &  0.9 &  1.33 &     \nodata   &AGN-2  &  3.0 & \\ 
601 &  03 32 18.46 & -27 45 56.0  &  1.4 &  0.73 &     \nodata   &AGN-2  &  3.0 & \\ 
602 &  03 32 22.00 & -27 46 56.1  &  1.2 &  0.67 &     \nodata   &AGN-2  &  3.0 & \\ 
603 &  03 32 57.69 & -27 47 10.9  &  1.5 &  2.04 &  1.56 -  2.32 &AGN-2  &  0.5 & \\ 
604 &  03 31 48.63 & -27 47 14.9  &  0.7 &  2.15 &  1.61 -  2.48 &AGN-2  &  0.5 & \\ 
605 &  03 32 39.17 & -27 48 32.4  &  1.5 &  4.29 &  4.13 -  4.35 &AGN-2  &  0.5 & \\ 
606 &  03 32 24.99 & -27 50 08.0  &  1.3 &  1.04 &     \nodata   &AGN-2  &  1.9 & \\ 
607 &  03 31 59.55 & -27 50 20.1  &  2.0 &  1.10 &  0.86 -  1.35 &AGN-2  &  0.6 & \\ 
608 &  03 33 03.86 & -27 50 26.3  &  1.7 &  0.89 &     \nodata   &AGN-2  &  3.0 & \\ 
609 &  03 32 36.19 & -27 50 37.1  &  1.0 &  1.86 &  1.69 -  2.02 &AGN-2  &  0.5 & \\ 
610 &  03 32 19.81 & -27 52 02.9  &  3.7 &  2.04 &  1.84 -  2.30 &AGN-2  &  0.5 & \\ 
611 &  03 32 41.56 & -27 43 27.9  &  2.7\tc  &  0.98 &     \nodata   &AGN-2  &  1.9 & \\ 
612 &  03 32 21.35 & -27 42 29.3  &  2.7 &  0.74 &     \nodata   &AGN-2  &  3.0 & \\ 
613 &  03 32 24.56 & -27 40 10.7  &  1.7 &  0.91 &  0.78 -  0.99 &AGN-2  &  0.9 & \\ 
614 &  03 32 34.82 & -27 40 42.1  &  2.6\tc  &  1.13 &  0.07 -  2.01 &AGN-2  &  0.5 & \\ 
615 &  03 32 01.30 & -27 50 50.9  &  1.5 &  0.76 &     \nodata   &AGN-2  &  3.0 & \\ 
617 &  03 32 31.43 & -27 57 26.4  &  1.6 &  0.58 &  0.51 -  0.64 &Galaxy  &  0.6 & \\ 
618 &  03 32 29.30 & -27 56 19.6  &  2.2 &  4.66 &  4.52 -  4.78 &AGN-1  &  0.5 & \\ 
619 &  03 31 55.63 & -27 54 02.4  &  1.1 &  1.94 &     \nodata   &AGN-1  &  3.0 & \\ 
620 &  03 32 30.18 & -27 53 06.1  &  1.8\tc  &  0.65 &     \nodata   &Galaxy  &  3.0 & \\ 
621 &  03 32 16.58 & -27 52 45.7  &  2.0 &  0.33 &  0.13 -  0.70 &Galaxy  &  0.5 & \\ 
622 &  03 32 50.00 & -27 44 07.3  &  3.6 &  1.75 &  1.39 -  2.11 &Galaxy  &  0.6 & \\ 
623 &  03 32 28.44 & -27 47 00.4  &  1.9\tc  &  1.74 &  1.55 -  1.94 &Galaxy  &  0.5 & \\ 
624 &  03 32 29.22 & -27 47 07.7  &  1.5 &  0.67 &     \nodata   &Galaxy  &  3.0 & \\ 
625 &  03 32 00.92 & -27 47 56.9  &  2.1 &  1.14 &  1.05 -  1.24 &Galaxy  &  0.6 & \\ 
626 &  03 32 09.47 & -27 47 57.2  &  1.7 &  1.90 &  1.74 -  1.95 &AGN-1  & 0.2\tablenotemark{k} & \\ 
627 &  03 32 23.36 & -27 48 52.6  &  2.5 &  0.25 &     \nodata   &Galaxy  &  3.0 & \\ 
628 &  03 32 55.20 & -27 51 02.8  &  4.2 &  2.07 &  1.95 -  2.27 &AGN-1  &  0.5 & \\ 
629 &  03 32 53.34 & -27 51 04.9  &  2.1 &  0.56 &  0.42 -  0.66 &Galaxy  &  0.5 & \\ 
630 &  03 32 28.29 & -27 44 03.5  &  0.6 &  3.25 &     \nodata   &AGN-1  &  3.0 & \\ 
631 &  03 32 15.08 & -27 43 35.5  &  1.7 &  1.40 &  0.76 -  1.55 &Galaxy  &  0.9 & \\ 
632 &  03 32 33.72 & -27 52 28.8  &  2.7 &  0.63 &  0.59 -  0.65 &AGN-2  &  0.5 & \\ 
633 &  03 31 50.43 & -27 52 12.3  &  1.7 &  1.37 &     \nodata   &AGN-2  &  3.0 & \\ 
634 &  03 32 51.46 & -27 47 47.3  &  1.2 &  1.40 &  1.28 -  1.54 &AGN-2  &  0.5 & \\ 
635 &  03 32 16.80 & -27 50 07.9  &  2.8\tc  &  0.73 &     \nodata   &AGN-2  &  2.0 & \\ 
636 &  03 31 50.24 & -27 50 46.9  &  4.5 &  0.80 &  0.70 -  1.32 &AGN-2  &  0.4 & \\ 
637 &  03 32 25.90 & -27 43 34.1  &  2.9\tc  &  0.76 &  0.65 -  0.99 &AGN-2  &  0.5 & \\ 
638 &  03 32 29.95 & -27 43 01.5  &  1.4\tc  &  1.39 &  1.29 -  1.42 &AGN-2  &  0.9 & \\ 
639 &  03 32 52.63 & -27 42 39.3  &  1.4 &  0.99 &  0.87 -  1.12 &AGN-2  &  0.6 & \\ 
641 &  03 32 39.17 & -27 59 19.2  &  1.4 &  0.74 &  0.67 -  0.82 &AGN-2  &  0.6 & \\ 
642 &  03 32 15.18 & -27 41 59.0  &  2.6 &  2.40 &     \nodata   &AGN-2  &  3.0 & \\ 
643 &  03 31 56.33 & -27 52 56.5  &  1.6 &  1.93 &  1.85 -  2.43 &Galaxy  &  0.5 & \\ 
644 &  03 32 45.96 & -27 57 45.6  &  1.2 &  0.12 &  0.10 -  0.15 &Galaxy  &  0.6 & \\ 
645 &  03 31 49.85 & -27 49 44.3  &  3.0 &  0.68 &     \nodata   &Galaxy  &  2.0 & \\ 
646 &  03 32 45.12 & -27 47 24.3  &  1.5 &  0.44 &     \nodata   &Galaxy  &  3.0 & \\ 
647 &  03 33 01.83 & -27 50 09.5  &  1.8 &\nodata&  \nodata &Star  &  3.0 & \\ 
648 &  03 32 46.54 & -27 57 13.4  &  2.7 &  0.77 &     \nodata   &Galaxy  &  3.0 & \\ 
650 &  03 33 07.33 & -27 44 32.9  &  0.9 &  0.21 &  0.17 -  0.26 &Galaxy  &  0.6 & \\ 
651 &  03 32 28.42 & -27 58 10.3  &  2.1 &  0.17 &  0.09 -  0.25 &Galaxy  &  0.6 & \\ 
652 &  03 32 49.33 & -27 43 02.4  &  1.1 &  0.08 &     \nodata   &Galaxy  &  3.0 & \\ 
653 &  03 33 03.72 & -27 44 11.0  &  0.9 &  0.91 &  0.66 -  1.15 &AGN-2  &  0.6 & \\ 
\enddata
\tablenotetext{a}{Unique CDFS identification number. See Table 2 of \citet{1ms}}
\tablenotetext{b}{Optical/infrared counterpart, in units of hr., min., sec., deg.,
min., and sec.} 
\tablenotetext{c}{Values with no range are spectroscopic redshifts 
(except stars)}
\tablenotetext{d}{At 95\% confidence level}
\tablenotetext{e}{0.2: HyperZ only; 0.3: BPZ only; 
0.4: COMBO-17 only; 0.5: BPZ and HyperZ; 0.6: COMBO-17 and HyperZ; 
0.7: COMBO-17 and BPZ; 0.9: COMBO-17, 
BPZ and HyperZ; 1.2: Single-line spectrum and HyperZ; 1.6: Single-line spectrum,
COMBO-17, and HyperZ; 1.9: Single-line spectrum, 
COMBO-17, BPZ and HyperZ ; 2.0: Secure spectroscopic redshift, but optical 
counterpart uncertain;  3.0: Secure spectroscopic redshift}
\tablenotetext{f}{Multiple sub-arcsecond structure in HST images}
\tablenotetext{g}{Based on power-law templates. BPZ value: $z = 1.33\ (1.02 - 2.54)$}
\tablenotetext{h}{Based on Hyperz. BPZ yields $z>7$}
\tablenotetext{i}{Based on the secondary value of BPZ. HyperZ: $z = 3.45 \ 
(3.10 - 3.86) $}
\tablenotetext{j}{Based on BPZ. HyperZ: $z = 3.66\  (2.95 - 5.12) $}
\tablenotetext{k}{Based on power-law templates. BPZ value: $z = 0.59\ (0.51 - 0.65)$}
\end{deluxetable}
}

\clearpage
{
\begin{deluxetable}{ccc}
\tablecaption{CDFS Sources without Redshift\label{tbl-2}}
\tablewidth{0pt}
\footnotesize
\tighttable
\tablehead{
\colhead{X-ray ID\tablenotemark{a}} & 
\colhead{$\alpha$(J2000)\tablenotemark{b}} & 
\colhead{$\delta$(J2000)\tablenotemark{b}} 
}
\startdata
261 &  03 31 57.11 & -27 51 10.9 \\
616 &  03 32 25.58 & -27 58 43.2 \\ 
640 &  03 32 17.74 & -27 38 52.1 \\
649 &  03 32 24.85 & -27 38 51.5 \\
\enddata
\tablenotetext{a}{Unique CDFS identification number}
\tablenotetext{b}{\ch\ coordinates, in units of hr., min., sec., deg., min., 
and sec. From \citet{1ms}} 
\end{deluxetable}
}
\clearpage

\end{document}